**Determinants of Phase-Separation Propensities, Material States, and Material Properties of Biomolecular Condensates**

Huan-Xiang Zhou*

Department of Chemistry and Department of Physics, University of Illinois Chicago, Chicago, IL 60607

*Correspondence: hzhou43@uic.edu

**Conspectus**

Phase separation of various materials has been studied for one and a half centuries. In the last two decades, phase separation of proteins and nucleic acids has received enormous attention, due its relevance to cellular functions. However, many of the observations on the resulting biomolecular condensates lack a theoretical underpinning. The first goal of this Account is to put forward theoretical frameworks for the phase-separation propensities, material states, and material properties of biomolecular condensates. Using these frameworks, I rationalize mechanistic interpretations from our recent experimental and computational studies, and synthesize these studies with prior literature to draw new conclusions. For phase-separation propensities, I relate the threshold (or saturation) concentration to the excess chemical potential in the dense phase, which in turn depends on intermolecular interaction strength and valency. For material states, I posit that liquid droplets form via complete phase separation, whereas amorphous dense liquids, reversible aggregates, and gels arise from premature termination of spinodal decomposition, due to overly weak or overly strong interactions or directional interactions. In particular, gels and aggregates are different forms of dynamically arrested states, with gels driven by tip growth via directional interactions whereas aggregates driven by monomer addition at interior sites to maximize valency. For material properties, I highlight the crucial roles of the stress relaxation time, which is determined by the mean lifetime of intermolecular bonds in a condensate. This relaxation time dictates how the condensate manifests viscoelasticity, including shear thickening and shear thinning, and accounts for the wide variation in zero-shear viscosity among different condensates. Validating and expanding these conclusions will further strengthen the theoretical foundations of the biomolecular condensate field. Other opportunities of particular interest include: bridging the gap between all-atom MD simulations and in vitro experiments, by extending the size and time limits of MD simulations via taking advantage of deep learning; expanding the experimental toolkit for the study of biomolecular condensates, by borrowing techniques from materials research; and bridging the gap between in vitro and in vivo studies, by testing mechanistic hypotheses inside cells.

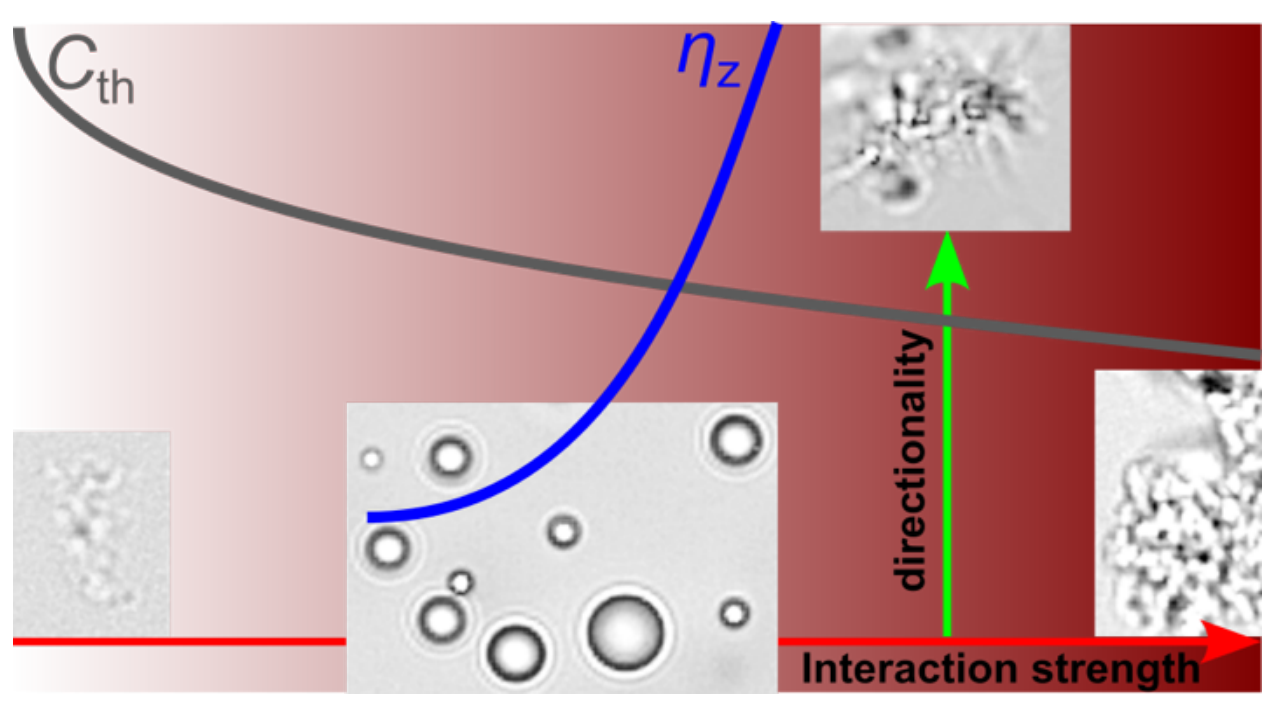

## 1. INTRODUCTION

Biomolecular condensates, formed via phase separation, exhibit a variety of material states.[1-6] Phase separation can occur spontaneously via spinodal decomposition, which begins with concentration fluctuations leading to dense regions and culminates in the formation of liquid droplets (Figure 1a).[7, 8] The latter appear to be the most common material state, inside which intermolecular interactions form and break (Figure 2a); droplets have a spherical shape and can fuse, both driven by interfacial tension. P granules, Cajal bodies, nucleoli, and stress granules possess these liquid-like properties.[9-12] I posit here that the other material states most probably result from premature termination of spinodal decomposition (Figure 1a), and hence correspond to dynamically arrested non-equilibrium states. Amorphous dense liquids (Figure 2b) have been reported recently to occur when overly weak intermolecular interactions prevent further condensation shortly after spinodal decomposition begins.[13] Transient clusters formed by the Mediator coactivator and RNA polymerase II are possible examples of this material state.[14] The third material state is gels, which emerge from arrested spinodal decomposition[15, 16] and can grow into system-spanning networks (Figure 2c). By deep quenching into the spinodal region, the N-terminal low-complexity region ($FUS_N$; residues 1-214) fused with a light-activatable domain (upon light activation)[17] and the peptide carboxybenzyl-capped di-phenylalanine (upon temperature decrease)[18] formed gels. The FG-rich domain of the yeast nucleoporin Nsp1p also formed gels; gelation is required for nuclear pore complex function.[19] P granules contain a peripheral gel phase that provides a stabilizing scaffold and helps with their localization in the posterior cytoplasm.[20] Cytoplasmic-abundant heat-soluble (CAHS) proteins in tardigrades undergo a gel transition to increase cell stiffness during dehydration.[21] Gels can be very porous and filled with a high level of water and thus are often referred to as hydrogels.[19, 21-25] The fourth material state is reversible aggregates, characterized by overly strong intermolecular interactions,[26] which prevent the completion of spinodal decomposition and produce particles that appear amorphous (Figure 2d).

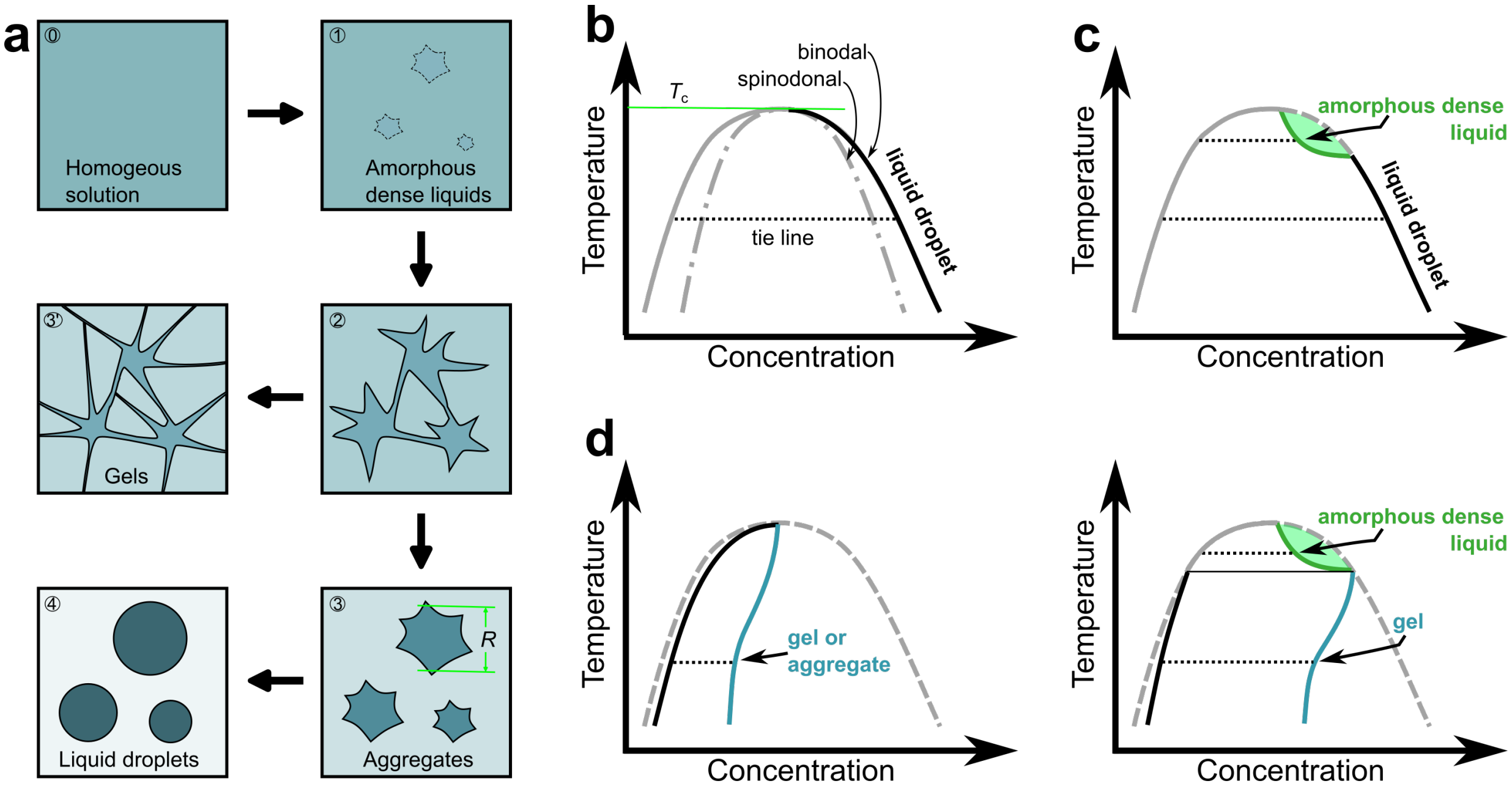

**Figure 1.** Phase separation and the resulting binodal. (a) Stages of phase separation via spinodal decomposition. (1) Concentration fluctuations in a supersaturation solution (box 0) produce dense regions; (2) condensation results in connected domains with multiple tips; (3′) tip growth yields system-spanning networks, whereas (3) further condensation toward domain centers breaks up inter-domain necks; (4) intra-domain rearrangement leads to droplets with smooth surfaces. Premature termination at stages (1), (3′), and (3) gives rise to amorphous dense liquids, gels, and aggregates, respectively. (b) The standard binodal and spinodal. (c) Binodal with right arm modified near the critical point due to formation of amorphous dense liquids. (d) Left: arrest line intersecting the left arm of the binodal. Here only gels or aggregates are possible in the dense phase; right: arrest line intersecting the right arm of the binodal. In the latter case, amorphous dense liquids or droplets can form at low quench and gels or aggregates can form at high quench. A solid tie line divides low and high quench.

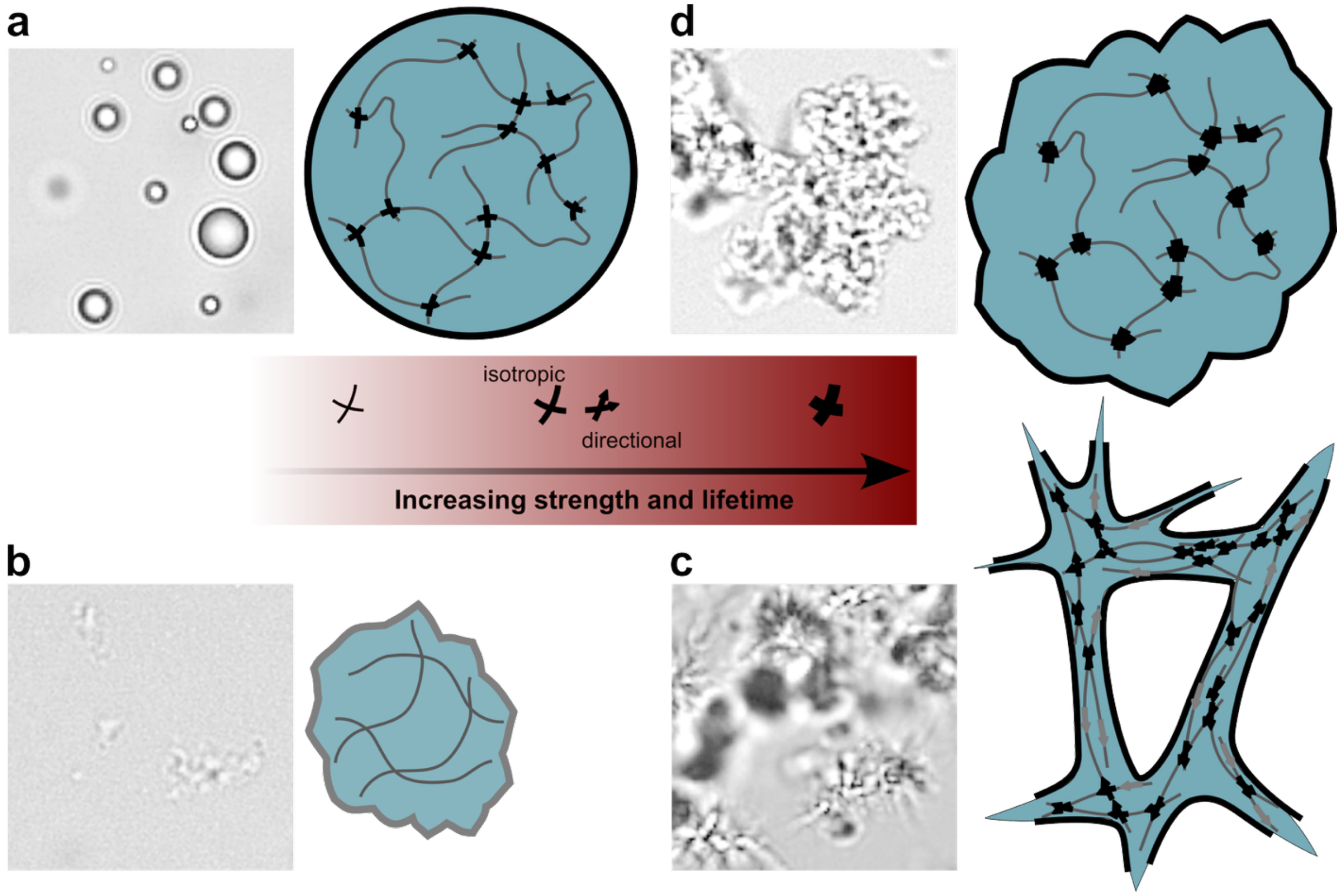

**Figure 2.** Four material states and their molecular organizations. (a-d) Brightfield images of (left) and illustration of intermolecular networks in (right) liquid droplets, amorphous dense liquids, gels, and reversible aggregates, respectively. Images are formed by XXssXX peptides, with X = L, A, I, and W, respectively; reproduced from ref 13. Intermolecular interactions, illustrated as occurring at chain crossings, are weak in amorphous dense liquids, moderate in droplets and gels, and strong in aggregates. Gels are special in that intermolecular interactions are directional, thereby guiding the growth and crosslinking of filaments.

While disordered proteins and RNAs are the main interest of current studies of biomolecular condensates, the earliest LLPS was observed on folded proteins, including arachin,[27] lysozyme,[28-34] γ-crystallins,[35-37] βB1-crystallin,[38] and bovine serum albumin.[34, 39] Lysozyme in particular has been a model system for observing dynamically arrested states including gels and aggregates.[15, 40, 41]

In addition to proteins, nucleic acids, and related biomolecules, a great variety of other materials undergo homotypic or heterotypic phase separation, including molecular fluids, polymers, colloids, and metallic alloys.[42-51] Accompanying experimental observations are theories that explain these observations and provide a conceptual framework. In particular, the PhD thesis of van der Waals[52] established the theoretical basis for the coexistence of gas and liquid phases for simple molecular fluids. Likewise, the Flory-Huggins theory[53, 54] laid the foundation for the phase-separation equilibrium of polymer solutions. The thermodynamic conditions for the coexistence of two phases (I and II) are that they have the same temperature, pressure, and chemical potential:

$$T^{\mathrm{I}} = T^{\mathrm{II}} \quad [1]$$

$$p^{I} = p^{II} \quad [2]$$

$$\mu^{I} = \mu^{II} \quad [3]$$

Biomolecular condensates are formed by biomolecules dissolved in water. For conceptual understanding, we will treat the main biomolecular components explicitly but water and other minor components (e.g., ions) implicitly, which in particular modulate the effective interactions of the explicit components. According to the Gibbs phase rule,[55] the number of independent intensive variables defining the boundary of two coexisting phases equals the number of components. Therefore, a one-component system can have two coexisting phases at various temperatures, up to a critical value $T_c$ (Figure 1b). The vapor and liquid phases of water provide a good example. The two phases differ in concentrations (or density in the case of water), except at the critical point; correspondingly, the binodal – i.e., the dependence of concentration on temperature – has two arms that meet at the critical point. Conditions [1]-[3] allow determination of the binodal (Figure 1b). For a two-component system, two phases can coexist at both variable temperatures and variable compositions (as measured by molar ratio). Equation [3] now expresses two equalities in chemical potential, one for each component. There are two extreme scenarios. In one, the two phases differ mainly in concentration, as in a one-component system. In the other, exemplified by an oil-vinegar mixture, the two phases differ mainly in composition, with one component dominant in the first phase and the other component dominant in the second phase. Most biomolecular condensates belong to the first extreme, so the vapor and liquid phases of water, not the separation of oil and vinegar, serve as the appropriate analogy.[8] However, in the dense phase of a multi-component system, the components can demix, leading to a multiphase organization where different regions are enriched in different components. This demixing within the dense phase has been called a second phase transition, with the first phase transition referring to the separation into the dilute and dense phases.[56]

Inside the binodal, the thermodynamically stable state is not a homogeneous solution, but two separated phases with different concentrations. This region can be further divided into a sub-region where the homogeneous solution is unstable, meaning that it will spontaneously phase separate via spinodal decomposition, and a sub-region where the homogeneous solution is metastable, and phase separation occurs via nucleation and growth. The boundary between the two sub-regions is called the spinodal, which also has two arms (Figure 1b). Theories for the kinetics of spinodal decomposition have also been developed,[57-59] which predict that the mean domain size ($R$; Figure 1a) is initially independent of time ($t$) and then grows according to a power law (i.e., $R \sim t^{\alpha}$).

In recent decades, computer simulations have played increasingly important roles in enhancing mechanistic understanding. A number of methods were developed to simulate phase separation.[7, 60-69] Some of these methods are based on continuum descriptions (phase-field models[64, 67, 68] and field-theoretic[63] simulations), but most are based on particle representations. These representations have become more realistic, from spherical particles and chains of identical beads in earlier days to coarse-grained protein or RNA chains and all-atom models of biomolecules in implicit or even explicit solvent at present.[13, 62, 70-87] Atomic details are extremely valuable for developing and validating mechanistic insights.[88]

This Account will present mechanistic insights into phase-separation propensities, material states, and material properties, derived from the integration of experimental and computational approaches in recent studies. On the experimental side, techniques include optical tweezers (OT)- and microscopy-based measurements of droplet fusion speed, interfacial tension, and viscosity.[89-91] On the computational side, techniques include all-atom explicit-solvent molecular dynamics (MD) simulations of phase separation[7] and atomistic modeling of protein-protein interactions and calculations of binodals.[62, 92, 93] Figure 3 lists the systems studied. They contain either a single component (Figure 3a, c) or two components (Figure 3b, d, e), corresponding to homotypic and heterotypic phase separation, respectively. The components include a small molecule adenosine triphosphate (ATP; Figure 3e), short peptides (Figure 3a), intrinsically disordered proteins (IDPs; Figure 3b, d, e), folded proteins (γ-crystallins; Figure 3c), a protein comprising five folded domains connected by flexible linkers (polySUMO; Figure 3b), and DNA (Figure 3d). The diversity of these systems ensures that the conclusions drawn below are general.

We got interested in ATP due to a suspicion that this small molecule, high on negative charges, might drive heterotypic phase separation with positively charged IDPs. We and others confirmed that this is indeed true, including with the small, Arg-rich IDP protamine (Figure 3e left).[94-100] The small sizes of ATP and protamine made it easy for us to run all-atom MD simulations of their heterotypic phase separation (Figure 3e right). We have continued interest in ATP's potential ability to tune condensate material properties. The basic design of the short peptides (to be denoted as XYssYX), with a disulfide bond as part of the backbone linker, was introduced by Spruijt and co-workers.[101] Such peptides formed droplets (XY = FF, LL, FL, and LF), aggregates (WW, FW, and WF), or gels (YF).[101, 102] Once again, the small sizes of the peptides enabled all-atom MD simulations of phase separation (Figure 3a right) and quantitative comparison with experimental measurements.[13, 87] The cellular function of protamine is to replace histones in the sperm nucleus to achieve extreme chromatin compaction.[103-106] We have used all-atom MD simulations (Figure 3d) and single-molecule force spectroscopy to uncover the mechanisms of protamine-mediated DNA condensation.[107-109] For γ-crystallins, our work[81] was motivated by the decades-long puzzle that paralogs with very high (>80%) sequence identities show large differences in the critical temperature for phase separation.[35, 37] For polySUMO and its partner molecule polySIM (10 linked SUMO-interacting motifs; Figure 3b), our interest was to understand how pH regulates the propensity of their heterotypic phase separation.[110]

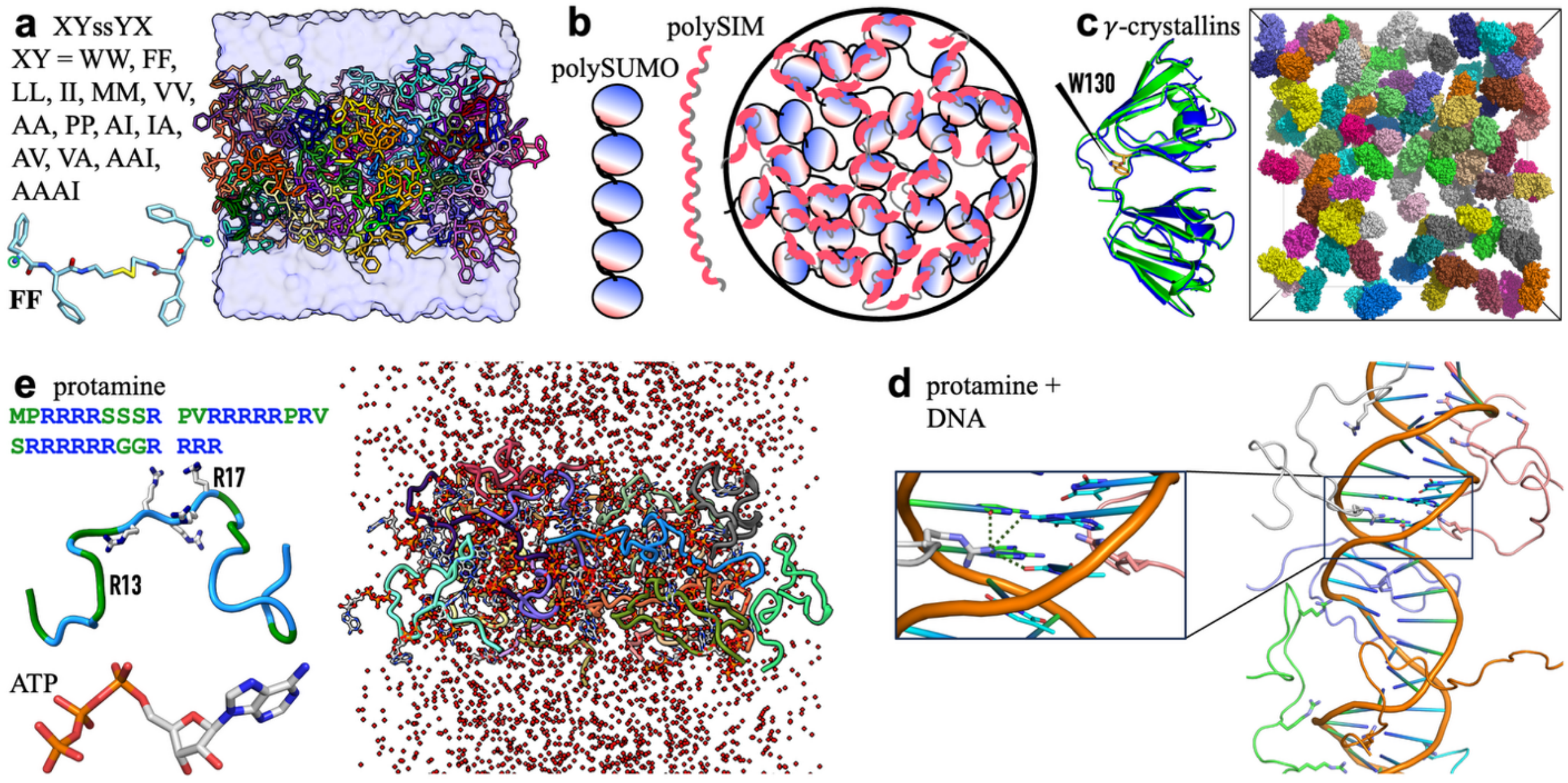

**Figure 3.** The main systems covered in this Account. (a) XYssYX peptides. The two titratable N-termini are highlighted by green circles. Adapted from ref 7. (b) polySUMO and polySIM. Adapted

from ref 110. (c) γ-Crystallins. Adapted from ref 81. (d) Protamine and DNA. Adapted from ref 107. (e) ATP and protamine. Adapted from ref 100.

## 2. PHASE-SEPARATION PROPENSITY IS GOVERNED BY INTERACTION STRENGTH AND VALENCY

Phase-separation propensities can be measured by the critical temperature $T_c$ (Figure 1b). However, $T_c$ is typically determined by fitting to the binodal, which requires concentration measurements in both phases over a range of temperatures. Concentration measurements in the dense phase are often difficult and hence $T_c$ is not routinely determined. Instead, a common measure for phase-separation propensity is the concentration, denoted as $C_{\text{th}}$, in the dilute phase at a fixed temperature. Because phase separation does not occur unless the concentration $C > C_{\text{th}}$, $C_{\text{th}}$ is the minimum, or threshold concentration required for phase separation. Results from Monte Carlo simulations of patchy particles have shown that the $T_c$-based measure of phase-separation propensity correlates very well with the $C_{\text{th}}$-based measure.[73] A low $C_{\text{th}}$ corresponds to a high phase-separation propensity.

To make use of Eq [3], note that the chemical potential can be decomposed into an ideal part and an excess part:

$$\mu = \mu_{\text{id}} + \mu_{\text{ex}} \quad [4a]$$

The ideal part is related to translation entropy and depends only on concentration, leading to

$$\mu = k_{\text{B}} T \ln(C/C_0) + \mu_{\text{ex}} \quad [4b]$$

where $k_{\text{B}}$ is Boltzmann's constant and $C_0$ is an unimportant constant. The excess part arises from interactions between solute molecules. We can neglect the excess chemical potential in the dilute phase, assuming that solute molecules there have little chance of encountering each other. Then Eq [3] becomes

$$-\ln(C_{\mathrm{th}}) = -\beta\mu_{\mathrm{ex}}^{\mathrm{II}} - \ln(C^{\mathrm{II}}) \qquad [5]$$

where $\beta = 1/k_{\mathrm{B}}T$ and the superscript II denotes the dense phase. Typically, for a given condensate, the dense-phase concentration ($C^{\mathrm{II}}$) is relatively constant for a given condensate. For example, for water, the liquid-phase density changes only 1.4-fold over the temperature from the triple point (0.01 °C) to 300 °C. In contrast, the vapor-phase density changes 9520-fold. If $C^{\mathrm{II}}$ is approximated as a constant, $C_{\mathrm{th}}$ is then determined by the excess chemical potential in the dense phase. $\beta\mu_{\mathrm{ex}}^{\mathrm{II}}$ can be expressed as[111]

$$e^{-\beta\mu_{\mathrm{ex}}^{\mathrm{II}}} =< e^{-U_{\mathrm{int}}^{\mathrm{II}}/k_{\mathrm{B}}T} >_{\mathrm{II}} \qquad [6]$$

where $U_{\mathrm{int}}^{\mathrm{II}}$ denotes the interaction energy of a solute molecule when inserted into the dense phase without perturbing its molecular networks, and $< \cdots >_{\mathrm{II}}$ means averaging over all positions of insertion inside the dense phase. Phase separation requires intermolecular interactions to be attractive (such that $\mu_{\mathrm{ex}}^{\mathrm{II}} < 0$ as demanded by $C_{\mathrm{th}} < C^{\mathrm{II}}$). Stronger intermolecular attraction in the dense phase results in a lower $\mu_{\mathrm{ex}}^{\mathrm{II}}$ and consequently a lower $C_{\mathrm{th}}$ and a higher phase-separation propensity.

Equation [5] qualitatively rationalizes many observations. For example, the increase in $C_{\mathrm{th}}$ with increasing temperature (left arm of the binodal in Figure 1b) is due to the fact that, in the expression of $\beta\mu_{\mathrm{ex}}^{\mathrm{II}}$ in Eq [6], the intermolecular attraction (i.e., $-U_{\mathrm{int}}^{\mathrm{II}}$) is scaled by $k_{\mathrm{B}}T$ and therefore raising temperature serves to weaken the effect of intermolecular attraction. Changes in other solvent conditions, such as pH or salt, that perturb the magnitude of intermolecular attraction will also shift $C_{\mathrm{th}}$. In particular, the XYssYX peptides have two N-termini (Figure 3a left, highlighted with green circles), which can change protonation states and hence electric charges as pH changes. Therefore, phase separation can be triggered by raising pH, which keeps the N-termini charge-neutral and hence minimizes net-charge repulsion between peptide chains.[13, 101] We measured the threshold concentrations of 8 nonpolar peptides (XXssXX, with X = W, F, L, M, I, V, A, P) as a function of pH (Figure 4a-d). For each peptide, as expected, $C_{\rm th}$ takes its lowest value at the highest pH. With decreasing pH, $C_{\rm th}$ increases gradually as the peptide N-termini become positively charged and net-charge repulsion intensifies. We anticipate that the $C_{\rm th}$ increase will continue up to a critical point, similar to a left arm of a typical binodal (Figure 1b).

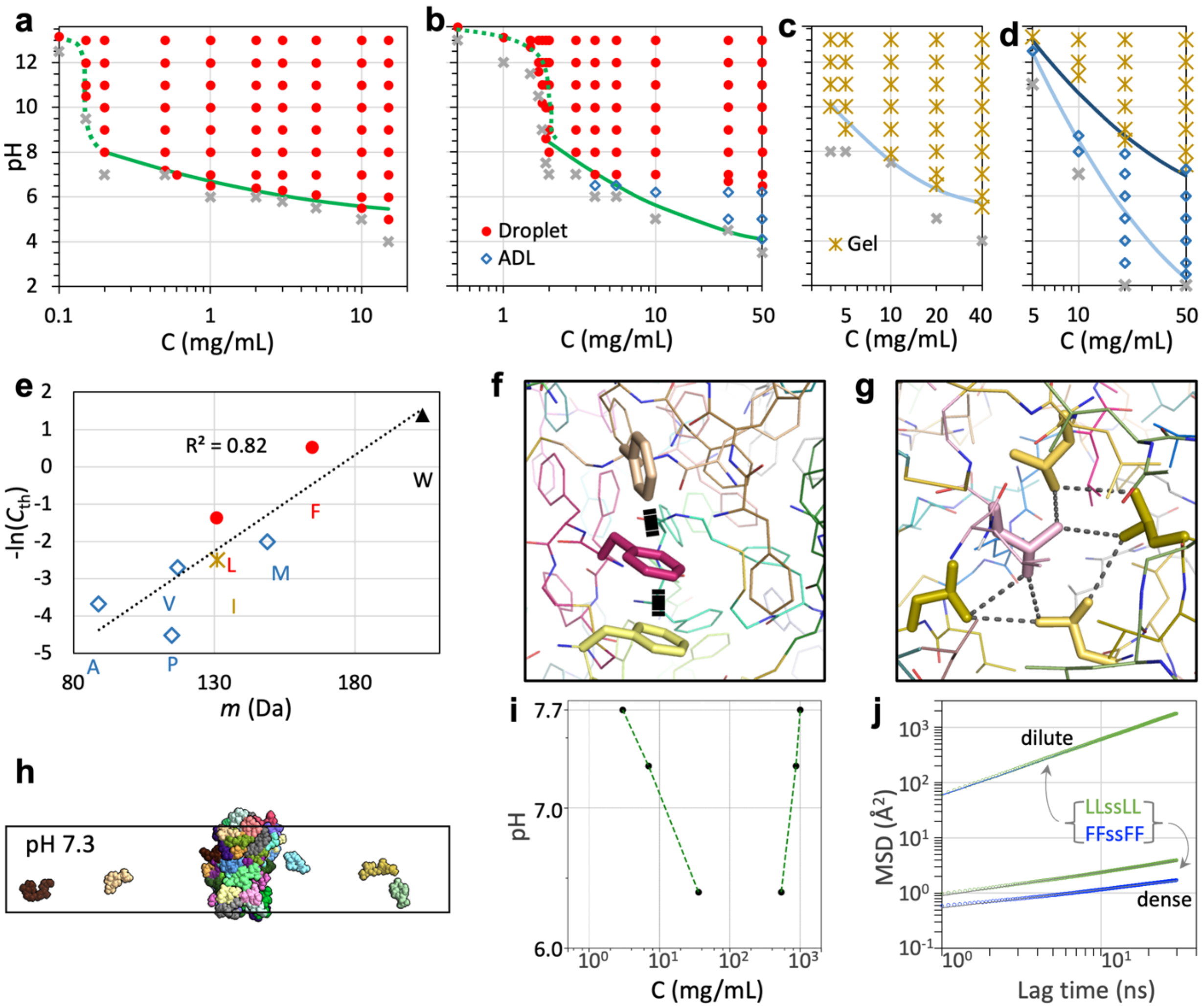

**Figure 4.** Phase-separation properties of XXssXX peptides. (a-d) Phase diagrams of FFssFF, LLssLL, 1:1 mixture of IIssII and AAssAA, and VVssVV. Similar to the mixture, IIssII only formed gels. (e) Correlation between threshold concentration at pH 7 and amino-acid molecular mass. (f, g) Molecular

organizations inside FFssFF (π-π stacks) and LLssLL (hydrophobic clusters) condensates. (h) Equilibration between dilute and dense phases in an LLssLL simulation in an elongated box. (i) Binodal calculated from simulations of (h). (j) Mean-square-displacements in the dilute and dense phases of (h). Adapted from ref 13.

At a fixed pH of 7, the threshold concentrations of the 8 peptides spread over a 368-fold range (Figure 4e). Interestingly, $-\ln(C_{\rm th})$ shows a strong correlation with the amino-acid molecular mass. The latter may be viewed as a surrogate of sidechain-sidechain interaction strength; hence the correlation is a manifestation of Eq [5]. Examination of the dense phase in phase-separation simulations (Figure 3a right) reveals extensive π-π interactions for WWssWW and FFssFF (Figure 4f) and hydrophobic clusters for the other nonpolar amino acids (Figure 4g). By extending the simulations in an elongated box, we observed equilibration between the two phases (Figure 4h), allowing for the calculation of the binodal (Figure 4i). For both FFssFF and LLssLL, the calculated $C_{\rm th}$ values follow the observed order (lower for FFssFF than for LLssLL), and are close to the experimental data. For example, the calculated $C_{\rm th}$ for LLssLL is 7 mg/mL, compared to the measured value of 4 mg/mL. However, for MMssMM, the calculated $C_{\rm th}$ is slightly lower than for LLssLL, contradicting the observed order (Figure 4e). We attribute this contradiction to the nonpolarizable force field used for the simulations. Nonpolarizable force fields tend to underestimate the magnitudes of the hydration free energies of sulfur-containing compounds,[112, 113] thereby making the interaction of the sulfur-containing M sidechain with water less favorable. Force-field reparameterization to correct this error is expected to reduce the drive for MMssMM phase separation, therefore leading to its correct order behind LLssLL. Indeed, phase-separation properties are extremely sensitive to force-field parameters and serve as excellent benchmarks for parameterization. The calculated binodals also validate the approximation regarding Eq [5], i.e., $C^{\rm II}$ is relatively constant. For LLssLL, on going from pH 7.7 to 6.4, $C_{\mathrm{th}}$ increases 12-fold while $C^{\mathrm{II}}$ decreases only 1.9-fold. The contrast is even more striking for FFssFF, with $C_{\mathrm{th}}$ increasing by > 65-fold but $C^{\mathrm{II}}$ decreasing by only 1.6-fold.

The ranking of phase-separation propensity obtained from short peptides, shown in Figure 4e, complements those derived from $C_{\mathrm{th}}$ changes by point mutations in IDPs.[26, 114-118] The latter results show that aromatic residues (W, Y, and F) and arginine (R) are the main drivers of homotypic phase separation of IDPs. These residues can form π-π and cation-π interactions. Interestingly, these residues are also the ones with high propensities to form local interactions within single IDP chains, as indicated by elevated NMR transverse relaxation rates of backbone amides.[119] Clusters of interaction-prone residues along an IDP's amino-acid sequence may play an outsized role in driving its phase separation. Such "key" motifs have been identified by NMR spectroscopy[120] and can be predicted by sequence-based methods.[121, 122]

In addition to pH, a change in salt concentration can also modulate the strengths of intermolecular interactions and hence phase-separation propensities.[117, 123-139] All-atom explicit-solvent simulations of the low-complexity domain of hnRNPA1 (A1-LCD) in a dense phase over a range of NaCl concentrations showed that the salt can have both direct and indirect effects on phase separation.[84] Diffuse ions in bulk water can screen charge-charge interactions between protein molecules. By binding to sidechain and backbone groups, ions can neutralize the IDP net charge (+9 for A1-LCD) as well as bridge between IDP chains. Additionally, salt ions can indirectly strengthen π-π, cation-π, and hydrophobic interactions by drawing water away from the interaction partners. Based on this understanding, we predicted four classes of salt dependence for phase-separation propensities using amino-acid composition. For IDPs that are high in charged residues but not in the net charge or aromatic residues, screening of favorable charge-charge interactions is the dominant salt effect, and consequently, phase-separation propensities decrease with increasing salt.[123, 125, 126, 128, 129, 131, 134, 137, 139] If, however, the aromatic content is also high, high salt can reinvigorate phase separation by indirectly strengthening π types of interactions, as in the case of FUS.[136] The designed IDP $(GRGDSPYS)_{40}$ is another example.[115] In contrast, A1-LCD represents IDPs that are high in both the net charge and aromatic content, a minimum salt concentration is required to screen out net-charge repulsion and trigger phase separation; higher salt then promotes phase separation.[137] The folded protein lysozyme behaves in the same way.[31]

Whereas pH regulates the homotypic phase separation of XXssXX peptides by changing their net charge and the resulting net-charge repulsion (Figure 4a-d), the effects of pH on the heterotypic phase separation of polySUMO and polySIM (Figure 3b left) are more subtle.[110] SUMO is a folded domain. A pH increase from 6 to 8 changes the charge on each of its three histidines (H) from +1 to 0. SUMO has a polarized electrostatic potential surface, with a negative face populated by D and E including E67 and a positive face where all three H residues are located. Charge neutralization of H reduces the intensity of the positive face. Consequently, both the heterotypic interaction with the negatively charged SIM and the homotypic interaction with the negative face of SUMO are weakened (Figure 3b right), as quantified by atomistic modeling of protein-protein interactions.

Compared to IDPs, our understanding on how sequence variations affect phase-separation propensities of folded proteins lags behind. A ~30 °C gap in $T_c$ between γ-crystallin paralogs with >80% sequence identities was reported four decades ago.[35] Yet it was only recently that we identified the substitution from an S in the low-$T_c$ γB to a W in the high-$T_c$ γF at residue 130 as a major contributor to the $T_c$ gap (Figure 3c left), based on atomistic modeling of protein-protein interactions and calculations of binodals (Figure 3c right).[81] As noted above, W is a main driver of homotypic phase separation of IDPs, so our identification of the S130W substitution is well-founded. However, while a W residue in an IDP can freely participate in $\pi$ types of interactions with other chains, this ability is curtailed in a folded protein. For example, a buried W cannot interact with other protein molecules. In the case of $\gamma$-crystallins, residue 130 is strategically located in the cleft between two domains (Figure 3c left). This location is well-suited for docking another domain and thus for protein-protein interactions, thereby amplifying the effect of the S130W substitution. Our binodal calculation yielded a $T_c$ of 4 °C for $\gamma$B, in agreement with the experimental value.[35, 37] The calculated $T_c$ for $\gamma$F was 10 °C, which, though moving in the right direction, is significantly lower than the experimental value of 39 °C.[37] There is uncertainty as to whether the latter value was partly due to multimerization via cysteine disulfide crosslinking, which occurs in hours[140] and raises $T_c$.[141] Multimerization increases valency, i.e., the number of interactions that a protein molecule can form simultaneously, and hence the phase-separation propensity. This rule applies to both folded proteins[34, 141, 142] and IDPs.[115]

While Eq [5] is the basis for the phase-separation propensities of both folded proteins and IDPs, the latter have several advantages.[8] First, every residue of an IDP can participate in intermolecular interactions, but only those on the surface of a folded protein can do so. Second, two IDP chains can readily adapt to each other to form multiple interactions, but doing so between two folded protein molecules requires shape complementarity. Third, an IDP chain, due to its open conformations, can have much higher valencies than a folded protein. For these reasons, IDPs typically have much lower threshold concentrations than folded proteins (~0.1 mg/mL vs. ~50 mg/mL[31, 37]).

## 3. DETERMINANTS OF MATERIAL STATES

In addition to the threshold concentrations of XXssXX peptides, Figure 4a-d also displays the material states formed under various combinations of peptide concentration ($C$) and pH.[13] FFssFF (Figure 4a) formed liquid droplets in all $C$–pH combinations tested; the same was true for LLssLL (Figures 4b and 2a) and MMssMM, except near the critical point, where amorphous dense liquids (ADLs) were formed. For PPssPP, the droplet region shrank to a sliver at high pH. IIssII (Figures 4c and 2c) formed gels in all $C$–pH combinations tested; VVssVV also formed gels, except near the critical point, where again ADLs were formed. AAssAA formed ADLs (Figure 2b) in almost the entire phase-separation region, except at the corner of high $C$ and high pH, where gels were formed. WWssWW formed reversible aggregates (Figure 2d) in all $C$–pH combinations tested. While the strength and valency of intermolecular interactions govern the threshold concentration for phase separation, the determinants of material states are more complex, as illustrated by the fact that LLssLL formed droplets but IIssII formed gels, despite the physicochemical similarity between L and I.

### 3.1. Do Material States Arise From Termination of Phase Separation at Different Stages?

I posit that liquid droplets form upon completion of phase separation, whereas the other material states arise from premature termination of spinodal decomposition (Figure 1a). As illustrated in MD simulations of simple particle and chain models, phase separation can occur spontaneously via spinodal decomposition.[7] Once a biomolecular solution is quenched inside the spinodal, by lowering temperature (Figure 1b) or raising pH in the case of XYssYX peptides, concentration fluctuations initially produce dense regions (Figure 1a box 1). Condensation then results in connected domains with multiple tips (Figure 1a box 2). If tip growth via monomer addition is favored, molecular networks will become system-spanning (Figure 1a box 3′). On the other hand, if monomer addition is favored at interior sites to maximize valency, then domains will be further condensed and inter-domain necks will be broken (Figure 1a box 3). Surface smoothing via molecular rearrangement finally leads to droplets (Figure 1a box 4). However, premature termination of spinodal decomposition at the stages represented by boxes 1, 3′, and 3 gives rise to ADLs, gels, and aggregates, respectively.

As FFssFF only formed liquid droplets, its boundary for the dense phase (Figure 4a) corresponds to the right arm of the standard binodal (Figure 1b). When the other material states occur, the dense-phase boundary needs to be modified. ADLs correspond to a carve-out starting at the critical point (Figure 1c). This carve-out is shallow for LLssLL (Figure 4b), but deeper for MMssMM and PPssPP. IIssII only formed gels (Figure 4c); its dense-phase boundary corresponds to a dynamic arrest line that intersects the left arm of the standard binodal (Figure 1d left panel). This arrest line also applies to WWssWW, which only formed reversible aggregates. VVssVV (Figure 4d) and AAssAA formed ADLs at low quench and gels at high quench. Correspondingly, the arrest line intersects the right arm of the standard binodal; the tie line at the intersection separates ADL formation and gel formation (Figure 1d right panel).

We performed all-atom MD simulations of XXssXX phase separation at high pH.[13] Starting from a homogeneous solution, FFssFF, LLssLL, and MMssMM condensed into a slab with relatively flat surfaces (Figure 3a right), matching with their observed droplet state (Figure 4a, b). AAssAA and PPssPP nearly condensed into a slab, but the surfaces were highly dynamic, consistent with the ADL state. IIssII and VVssVV condensates had holes, resembling gels. WWssWW condensates were densely packed and irregularly shaped, befitting the observed aggregate state. Below, I further compare and contrast the four material states.

### 3.2. Amorphous Dense Liquids: A New Material State

ADL samples are transparent to the naked eye; brightfield images show very low contrast with cloud-like morphology (Figure 2b). These features suggest a low packing density, in keeping with ADL's location near the critical point (Figure 1c). Our 2024 paper[13] is the first to report the ADL phase for biomolecular condensates. Small transient clusters formed by the Mediator coactivator and RNA polymerase II[14] share two important features with ADLs: very small dense cores (Figure 2b left) and highly dynamic nature (according to MD simulations[13]), making them possible examples of ADLs. We borrowed the term "dense liquid" from the field of crystal growth, where it refers to a form of precursors to crystal nucleation.[143] "Amorphous" describes the irregular shapes of the dense regions (Figure 2b), not the disordered nature of molecular arrangements inside, as that should be taken for granted for any liquid. ADLs may be viewed as precursors to liquid droplets if phase separation can proceed to completion. Note that, even though we use the descriptor "dense", ADLs are actually less concentrated than the corresponding liquid droplets, just like dense-liquid precursors are less concentrated than the corresponding crystals.[143]

That ADLs are precursors to liquid droplets is supported by the presence of droplets in the midst of ADLs, in particular in the transition zone between the pure ADL subregion and the pure droplet subregion for XXssXX with X = L, M, and P. The failure of ADLs to further condense into liquid droplets can be attributed to overly weak intermolecular interactions. Indeed, the location of ADLs near the critical point (Figures 1c and 4b, d) means that any further weakening of interactions would push the system beyond the critical point, where phase separation could no longer happen. The fraction of the phase-separation region covered by ADL is also telling. For LLssLL, ADL occupied only a small fraction of the entire phase-separation region (Figure 4b). This fraction expanded to slightly higher pH for MMsMM and to even higher pH for VVssVV (Figure 4d). For PPssPP and AAssAA, ADLs covered most of the phase-separation region. The ADL area fraction thus expanded as the intrinsic interaction strength, as reported by $C_{\mathrm{th}}$ (Figure 4e), weakened.

Let me reframe the occurrence of ADLs in thermodynamic instead of molecular terms. At the critical point, the dilute and dense phases lose distinction. Just below the critical point, the thermodynamic drive for phase separation approaches zero; correspondingly, phase separation proceeds exceedingly slowly, a phenomenon known as critically slowing down.[144] Therefore, ADLs are kinetically trapped. This conclusion was validated by heating ADL samples.[13] When the temperature was elevated from 25 °C to 100 °C, the droplet population increased. The elevated temperature facilitates overcoming kinetic barriers, even as it reduces the thermodynamic drive for phase separation.

### 3.3. Aggregates and Gels: Dynamically Arrested States

Compared to droplets, aggregates are more densely packed, as indicated by ready sedimentation and very high contrast in brightfield images (Figure 2d).[13] XYssYX gels, on the other hand, are marked by system-spanning networks, which stick to container walls, resulting in a loss of fluidity. Brightfield images show filaments emanating from dense centers, resembling sea urchins.[13, 87, 102] Both aggregates and gels are solid-like, in that molecules are largely immobile as if locked in interaction networks, as indicated by fluorescence recovery after photobleaching (FRAP).[13, 87] The solid-like behavior of aggregates can be attributed to overly strong intermolecular interactions, consistent with high packing density. The nature of intermolecular interactions driving gel formation is more complex and will be delineated below.

Gels have been reported to emerge from arrested spinodal decomposition.[15, 16] Using phase contrast microscopy, Cardinaux et al.[15] observed the completion of phase separation upon a low quench of lysozyme solutions but arrested spinodal decomposition at high quench. The contrasting behaviors at low and high quenches are similar to those seen in MD simulations of simple particle and chain models,[7] and lead to an arrest line intersecting with the right arm of the binodal (Figure 1d right panel). For lysozyme, Cardinaux et al. determined this arrest line by measuring the protein concentration in the gel phase after centrifugation.

Relative to gel formation, the mechanism of aggregate formation in relation to phase separation has received little attention. In their study of six folded proteins, Dumetz et al.[40] interpreted the formation of gels and reversible aggregates "as two manifestations of the same phenomenon" (i.e., dynamic arrest). That is the starting point of my posit in Subsection 3.1 regarding gels from aggregates, but the next question is: what distinguishes these two material states? Lattice simulations by Kallala et al.[145] provided a clue. Their simulations produced dense particles when monomer addition was favored at sites with high preexisting coordination numbers, but branched polymers when tip growth was favored. These results served as inspiration for my conclusion that aggregates are driven by monomer addition at interior sites to maximize valency (Figure 1a box 3), whereas gels are driven by tip growth, in particular via directional interactions (Figure 1a box 3′). These points are elaborated next.

### 3.4. Interaction Strength Is a Major Determinant of Material States, but Not the Only One

Subsection 3.2 states that ADLs are due to overly weak intermolecular interactions. "Overly weak" means weaker than the counterparts in droplets. The validity of this point is evidenced by the ADL subregion being at lower pH than the droplet subregion (Figure 4b); lowering pH increases net-charge repulsion and thus weakens overall intermolecular attraction. By the same token, intermolecular interactions in ADLs are also weaker than in gels (Figure 4d). Similarly, by adding urea to reduce peptide backbone hydrogen bonding and thus overall intermolecular attraction, the material state of VVssVV changed from gels to ADLs.[13]

By contrast, Subsection 3.3 attributes aggregates to overly strong intermolecular interactions. This point is validated by the transitions in the material state of a polylysine–heparin condensate as salt was added to screen heterotypic electrostatic attraction.[89] The condensate was reversible aggregates below 0.8 M KCl, liquid droplets at higher KCl, and dissolved above 1.2 M KCl. We observed similar results with an IDP-DNA condensate (unpublished). Protamine–DNA mixtures formed aggregates; in OT-based single-molecule studies, protamine condensed DNA into hyper-stable "tangles".[107] Our all-atom MD simulations suggested that hyperstability stems from the strong tendency of protamine Arg sidechains to form hydrogen bonds simultaneously with multiple nucleobases, frequently adopting a wedge configuration between the two strands of DNA (Figure 3d). Interestingly, with single-stranded (ss) instead of double-stranded (ds) DNA, the condensate became liquid droplets; concomitantly, the stability of protamine-condensed ssDNA structures was moderated in single-molecule studies.[108] The reduced stability can be explained by the change in Arg-nucleobase interactions from wedges to cation–π. HMGB1, a protein with two folded domains and a disordered tail, formed droplets with dsDNA, suggesting weaker heterotypic interactions than those mediated by protamine.[109] As corroboration, adding HMGB1 converted protamine-dsDNA aggregates into droplets. The shift in material states may have physiological relevance during spermiogenesis, when histones are replaced by protamine.

So there is a clear order in interaction strength between ADLs, droplets, and aggregates, from weak to strong. What about gels? Interactions are stronger in gels than in ADLs as noted above, and possibly even stronger than in droplets. For example, adding weak-interaction AAssAA converted AIssIA gels into droplets.[87] However, interaction strength is not the only determinant for gel formation, as illustrated by the observation that IIssII formed gels but LLssLL formed droplets[13] and even more intriguing observation that AIssIA formed gels but IAssIA formed droplets.[87] For clues, we turned to all-atom MD simulations.[13, 87] They suggested backbone hydrogen bonding as a possible determinant for gel formation. For example, $1.0 \pm 0.1$ hydrogen bonds per half chain were formed in the IIssII condensate, but this number reduced to $0.67 \pm 0.09$ in the LLssLL condensate.

### 3.5. Backbone Hydrogen Bonding Promotes Gel Formation via Tip Growth

To verify that backbone hydrogen bonding is indeed a determinant of gel formation, we reduced hydrogen bonding by various means and assessed whether each disrupted gel formation of AIssIA.[87] These included adding urea or the low-hydrogen-bonding peptide AAssAA, extending the sequence at both termini with Ala residues, and methylation of a backbone amide. The outcomes clearly support the importance of backbone hydrogen bonding in gel formation. AIssIA gels were either dissolved (by adding urea) or converted to droplets (by adding AAssAA, methylating outer amides, or extending the sequence to AAAIssIAAA) or ADLs (by methylating inner amides).

How does backbone hydrogen bonding promote gel formation? Our all-atom MD simulations revealed that peptide chains that engage in hydrogen bonding tend to be straight, i.e., with the two halves directed away from each other. The straight conformation allows molecular networks to grow in a fixed direction, reminiscent of the tip growth in the lattice simulations of Kallala et al.[145] Tip growth is supported by the sea-urchin morphology of gels (Figure 2c). The sea-urchin morphology has been observed in lysozyme crystallization and interpreted as directional growth from a polycrystalline center.[31]

So now it is clear that, while LLssLL formed droplets, IIssII formed gels due to its higher tendency for backbone hydrogen bonding, but what is the origin of this higher tendency? We have traced the origin to β-branching.[87] It is well-known that β-branched amino acids stabilize β-sheets by enhancing their rigidity. Hence VVssVV also formed gels (Figure 4d) as Val is also β-branched. We can also explain why AIssIA formed gels but IAssIA formed droplets by noting that the rigification effect of β-branching starts at the second (inner for XYssYX) position of β-strands. As confirmation, AVssVA formed gels but VAssAV formed droplets.

We used a multitude of experimental techniques to distinguish the structural, dynamic, and material properties AIssIA gels and IAssAI droplets, including FRAP, dynamic light scattering, NMR relaxation, Fourier-transform infrared spectroscopy (FTIR), transmission electron microscopy (TEM), and $^{1}$H-$^{1}$H nuclear Overhauser effect spectroscopy (NOESY). With gel samples, FTIR confirmed backbone hydrogen bonding by an amide I′ peak at 1625 cm$^{-1}$; TEM images showed highly cross-linked filaments ~10 nm width; $^{1}$H-$^{1}$H NOESY, coupled with MD simulations, revealed strong, backbone hydrogen bonding-buttressed molecular networks.

## 4. OBSERVATIONS ON MATERIAL PROPERTIES

For condensates in the liquid-droplet state, we have used OT and microscopy to measure a set of material properties comprising interfacial tension, fusion speed, and viscoelasticity.[89-91] Given the complex molecular networks and intermolecular interactions (Figure 3) and their dynamics at a wide-range of spatial and temporal scales, molecular-level interpretations of these properties should be seen as tentative.

### 4.1. Why Do Biomolecular Condensates Have Similar Surface Tensions?

Surface tension is a fundamental property of all liquids, including biomolecular condensates in the liquid-droplet state. Indeed, the spherical shape of liquid droplets is a direct manifestation of surface tension, which drives the minimization of surface area. Our 2024 review[8] tabulated 39 values of condensate surface tension from direct measurements; most of them fall into a narrow range of 20 to 200 pN/μm. Our surface tension measurements involve stretching a droplet using two optically trapped microbeads.[90, 146] The results for ATP-IDP condensates were 24 pN/μm when the IDP was protamine and 27-60 pN/μm when the IDP was polylysine.[100] For XXssXX peptide condensates, the values were 96, 109, and 38 for X = F, L, and M, respectively.[13] These measurements reinforce the narrow range of surface tension.

As the simplest model, we consider solute molecules modeled as spherical particles interacting with a centrosymmetric potential $u(r)$. The surface tension of this model (away from the critical point) is[147, 148]

$$\gamma = \frac{\pi {C_{\mathrm{II}}}^2}{8} \int_0^\infty r^4 \frac{du(r)}{dr} g(r) dr \qquad [7a]$$

$$= \frac{9{\phi_{II}}^2 I}{2\pi} \frac{k_{\mathrm{B}}T}{d^2} \qquad [7b]$$

where $d$ is the hard-core diameter of the particles, $g(r)$ is the radial distribution function, $\phi_{II} = (\pi/6)\sigma^3 C_{\mathrm{II}}$ is the solute volume fraction, and

$$I = \int_0^{\infty} x^4 \frac{d\beta u(dx)}{dx} g(dx)dx \quad [8a]$$

For a square-well potential with range $[d, \lambda d]$, this integral is

$$I = \lambda^4[g(\lambda^+ d) - g(\lambda^+ d)] - g(d^+) \quad [8b]$$

Aarts et al.[149] noted that the scaling relation in Eq [7b],

$$\gamma \sim \frac{k_{\mathrm{B}}T}{d^2} \quad [9]$$

explains why the surface tension of phase-separated polymer-colloid mixtures is orders of magnitude lower than that of small-molecule fluids like water ($\gamma$ = 72 × $10^3$ pN/μm at room temperature), due to the much larger size of colloid particles. The same argument applies to biomolecular condensates. This scaling relation also explains why condensate surface tensions fall into a narrow range, since biomacromolecules have roughly similar sizes (a few nm).

Just like $T_{\mathrm{c}}$, $\gamma$ increases with increasing attraction (see Eq [7a]). From molecular simulations, we have observed that perturbations (e.g., by adding another component) that change $T_{\mathrm{c}}$ also change $\gamma$ in the same direction.[78] This correlation was confirmed in an experimental study, where crowding agents were found to increase both phase-separation propensity and surface tension.[91]

### 4.2. The Different Facets of Condensate Viscoelasticity

It is now well recognized that biomolecular condensates are viscoelastic materials, rather than purely viscous liquids. Viscous and elastic moduli have been measured using OT-trapped microbeads, in either an active[90] or passive[150] mode. These moduli are the real and imaginary parts, respectively, of the complex shear modulus $G^*(\omega)$, which in turn is the constant of proportionality between shear stress and shear strain at a given angular frequency $\omega$. We have fit data to standard viscoelastic models, in particular the Bergers model,[90, 91, 100, 151]

$$G^*(\omega) = \frac{i\omega\eta_0}{1 + i\omega\tau_0} + \frac{i\omega\eta_1}{1 + i\omega\tau_1} \tag{10}$$

The corresponding complex viscosity, relating shear stress to shear rate, is

$$\eta^*(\omega) \equiv \frac{G^*(\omega)}{i\omega} = \frac{\eta_0}{1 + i\omega\tau_0} + \frac{\eta_1}{1 + i\omega\tau_1} \tag{11}$$

The complex viscosity is the Fourier transform of the shear relaxation modulus, which in the present case is

$$G(t) = \frac{\eta_0}{\tau_0} e^{-t/\tau_0} + \frac{\eta_1}{\tau_1} e^{-t/\tau_1} \tag{12}$$

Each term on the right-hand side is a Maxwell component. In polymer solutions, the longer relaxation time $\tau_1$ is usually identified as the chain reconfiguration time; for associative polymers, which may be taken as a model for biomolecular condensates, $\tau_1$ is determined by the mean lifetime of interchain crosslinks (see Eq [23b′] below). The measured values of $\tau_1$ range from a few ms to 1 s.[90, 91, 100, 151] When the stress relaxation time ($\tau_0$ or $\tau_1$) $\rightarrow 0$, the Maxwell component becomes Newtonian (or purely viscous). For processes occurring on timescales much longer than $\tau_1$, the condensate will effectively behave as purely viscous; the relevant viscosity is

$$\eta_{\mathrm{z}} \equiv \lim_{\omega \to 0} \eta^*(\omega) = \eta_0 + \eta_1 \tag{13}$$

where the subscript "z" means "zero-shear". For processes on the same timescale as $\tau_1$ (but faster than $\tau_0$), elasticity will play a role, which is measured by the plateau modulus

$$G_1 \equiv G(t)|_{\tau_0 \ll t \ll \tau_1} = \frac{\eta_1}{\tau_1} \tag{14}$$

*4.2a Oscillatory vs. steady shear.* Instead of oscillatory shear, viscoelastic materials can also be characterized at a constant shear rate ($\dot{\sigma}$). At low shear rates, the viscosity is $\eta_{\mathrm{z}}$ of Eq [13]. However, at higher shear rates, viscosity becomes $\dot{\sigma}$-dependent; increases and decreases are known as shear thickening and thinning, respectively. Shear thinning is common in polymer solutions and is usually attributed to the alignment of polymer chains when they are disentangled and stretched by the shear flow. Shear thickening is sometimes observed in associative polymers, and is explained as due to flow-induced strengthening of intermolecular networks.

There is a long history of relating the complex viscosity $\eta^*(\omega)$ to the steady-shear viscosity $\eta(\dot{\sigma})$. Based on empirical observations, Cox and Merz[152] proposed the following identity between the magnitude of $\eta^*(\omega)$ and the steady-shear viscosity $\eta(\dot{\sigma})$:

$$|\eta^*(\omega)| = \eta(\dot{\sigma})|_{\dot{\sigma}=\omega} \qquad [15]$$

Cross introduced a model for shear thinning[153]

$$\frac{\eta(\dot{\sigma}) - \eta(\infty)}{\eta_{\mathrm{z}} - \eta(\infty)} = \frac{1}{1 + (\tau_{\mathrm{c}}\dot{\sigma})^m} \qquad [16]$$

and subsequently proposed the identity[154]

$$\tau_{\mathrm{c}} = \frac{\eta_{\mathrm{z}}}{G_1} \qquad [17]$$

Comparison with Eq [14] suggests that $\tau_{\mathrm{c}}$ can be identified as the shear relaxation time $\tau_1$.

*4.2b Effective viscosity depends on timescales.* In essence, $\eta^*(\omega)$ is the measure of viscosity at different frequencies or, equivalently, at different timescales. We have taken the explicit view that the "effective" viscosity, $\eta_{\mathrm{eff}}$, of a viscoelastic material depends on the dynamic process of interest, in particular, its timescale.[90] As stated above, for processes longer than $\tau_1$, the effective viscosity is $\eta_{\mathrm{z}}$. For faster processes, $\eta_{\mathrm{eff}}$ will be different from $\eta_{\mathrm{z}}$. We borrow the terminology shear thinning and thickening to describe cases where $\eta_{\text{eff}} < \eta_{\text{z}}$ and $\eta_{\text{eff}} > \eta_{\text{z}}$, respectively. For example, for a process occurring on a timescale much longer than $\tau_0$ but much shorter than $\tau_1$, the first Maxwell component is effectively Newtonian but the second Maxwell component may not be "activated", the effective viscosity would be $\eta_0$, and shear thinning would be manifested.

A related phenomenon, namely the dependence of effective condensate viscosity on spatial scales, has recently been reported.[79] These authors measured the diffusion constants of inert probes of varying sizes inside condensates and obtained the corresponding effective viscosity using the Stokes-Einstein equation. For probes larger than 100 nm, the effective viscosity is the macroscopic viscosity; for probes smaller than the sizes of the condensate-forming IDPs, the effective viscosity approaches that of the solvent. Interpreting their results using Eq [11], the diffusion of large probes requires the reconfiguration of IDP chains, hence the effective viscosity is $\eta_{\text{z}}$. In contrast, the diffusion of small probes does not perturb chain configurations and hence does not activate the second Maxwell component; consequently, the effective viscosity is $\eta_0$.

*4.2c Shape recovery of viscoelastic droplets.* When a liquid droplet is deformed from a spherical shape, surface tension drives it back into the spherical shape. I now use shape recovery as a concrete example of dynamic processes to illustrate the effects of viscoelasticity. This example is particularly useful because an analytical solution has been obtained.[155] The analytical solution has been validated by finite-element calculations.[156]

When a purely viscous droplet is deformed to a shape represented by a Legendre polynomial of order $l$ with amplitude $f_l(0)$, the recovery of the spherical shape is an exponential function of time. The Laplace transform of the deformation amplitude is

$$\hat{f}_l(s) = \frac{f_l(0)}{s + \lambda_l^{\mathrm{D}}} \quad [18a]$$

The rate constant is

$$\lambda_l^{\mathrm{D}} = \frac{l(l+2)(2l+1)}{2(2l^2+4l+3)} \frac{\gamma}{\eta_{\mathrm{z}} R} \quad [18b]$$

Note that $\lambda_l^{\mathrm{D}}$ increases with $\gamma$ but decreases with $\eta_{\mathrm{z}}$. For a viscoelastic droplet, shape recovery is no longer a single exponential; the deformation amplitude is now

$$\hat{f}_l(s) = \frac{f_l(0)}{s + \lambda_l^{\mathrm{D}}/\hat{g}(s)} \quad [19a]$$

with

$$\hat{g}(s) = \frac{\eta^*(\omega)|_{i\omega=s}}{\eta_0} \quad [19b]$$

For a Jeffreys model of viscoelasticity (Eq [11] with $\tau_0 \to 0$), shape recovery is bi-exponential, with amplitudes $A_{l\pm}$ that add up to 1 and rate constants $\lambda_{l\pm}^{\mathrm{D}}$ that are higher and lower than the purely viscous counterpart, $\lambda_l^{\mathrm{D}}$, respectively.

When stress relaxation is fast relative to the shape recovery of a purely viscous droplet (i.e., $\lambda_l^{\mathrm{D}}\tau_1 \ll 1$), the "–" mode dominates (i.e., $A_{l-} \gg A_{l+}$), and the effective rate constant is $\lambda_{l-}^{\mathrm{D}}$. Because $\lambda_{l-}^{\mathrm{D}} < \lambda_l^{\mathrm{D}}$, the effective viscosity is higher than $\eta_{\mathrm{z}}$, indicating shear thickening, although the increase in viscosity [$\approx (\lambda_l^{\mathrm{D}}\tau_1)\eta_1$] is small. In contrast, when stress relaxation is slow relative to the shape recovery of a purely viscous droplet (i.e., $\lambda_l^{\mathrm{D}}\tau_1 \gg 1$), the "+" mode dominates (i.e., $A_{l+} \gg A_{l-}$). Now $\lambda_{l+}^{\mathrm{D}}$ is the effective rate constant and shear thinning is expected. In fact, $\lambda_{l+}^{\mathrm{D}}$ approaches $(\eta_{\mathrm{z}}/\eta_0)\lambda_l^{\mathrm{D}}$, which is the rate constant for a viscous droplet with viscosity $\eta_0$. This limit exemplifies the aforementioned scenario where the Maxwell component with relaxation time $\tau_1$ is "activated". The level of shear thinning can be very large since $\eta_0$ can be significantly smaller than $\eta_z$, as shown next.

*4.2d Shear thickening and thinning in droplet fusion.* The effects of viscoelasticity on the dynamics of droplet shape recovery and fusion are qualitatively similar.[156] Indeed, the predictions of the analytic solution for shape recovery are consistent with our experimental observations, including the manifestation of both shear thickening and thinning[90] and significant levels of shear thinning.[13, 100]

For the fusion of two equal-sized Newtonian (i.e., purely viscous) droplets, we fit the finite-element numerical solution for the fusion progress curve, as monitored by the edge-to-edge distance, to a stretched exponential, $1 - e^{-(t/\tau_{fu}^{\mathrm{N}})^{1.5}}$, with the fusion time given by[89]

$$\tau_{fu}^{\mathrm{N}} = \frac{1.97\eta_{\mathrm{z}}R}{\gamma} \qquad [20]$$

where $R$ is the initial droplet radius. The scaling relation of fusion time with $\eta_{\mathrm{z}}R/\gamma$ is widely used in the condensate community. Our OT-directed fusion progress curves all fit well to the stretched exponential function.[13, 89-91, 100] We then use the resulting fusion time, $\tau_{\mathrm{fu}}$, to define an effective viscosity

$$\tau_{\mathrm{fu}} = \frac{1.97\eta_{\mathrm{eff}}R}{\gamma} \qquad [21]$$

By using OT to measure all three properties: surface tension, viscoelasticity, and fusion time on the same condensates, we observed the breakdown of the scaling relation of Eq [20] for the first time.[90] Shear thickening was observed on condensates where stress relaxation time $\tau_1$ was shorter than the fusion time $\tau_{\mathrm{fu}}$, whereas shear thinning was observed on condensates with $\tau_1 >$ $\tau_{\mathrm{fu}}$, consistent with the analytic solution for shape recovery. The levels of deviations, though, were relatively small (3.2-fold for thickening and 1.2-fold for thinning).

In our recent studies of ATP-protamine condensates, we observed 116-fold shear thinning.[100] According to the analytic solution of shape recovery, significant shear thinning occurs when $\tau_1 \gg \tau_{\mathrm{fu}}$ and $\eta_0 \ll \eta_{\mathrm{z}}$. Both conditions were indeed met: $\tau_1$ was ~1000 ms whereas $\tau_{\mathrm{fu}}$ was ~40 ms; $\eta_0$ was ~0.01 Pa s whereas $\eta_{\mathrm{z}}$ was ~16 Pa s. For molecular interpretation, we attributed the fast fusion (relative to a viscous model) and low $\eta_0$ (relative to $\eta_{\mathrm{z}}$) to the small size of ATP, which can readily adapt to the configurations of protamine chains. The long $\tau_1$ may be attributed to the slow reconfiguration of protamine chains (see Eq [23b′] below) because of their strong interaction with ATP (Figure 3e). We also observed significant shear thinning on XXssXX peptide condensates,[13] ranging from 7- and 85-fold for X = F, L, and M. Here again the small size of the peptides may be a factor.

### 4.3. Why Do Condensate Viscosities Vary So Much?

The large size of biomacromolecules was used above to explain why the surface tension of their condensates is orders of magnitude lower than that of small-molecule fluids like water. The viscosities of biomolecular condensates (0.1 to 1000 Pa·s) are orders of magnitude higher than that of water (0.001 Pa·s).[8] Again, molecular size is a major factor. This point can be illustrated by a simple model of polymer solutions, comprising independent Rouse chains, each with *N* beads connected by springs.[157] The zero-shear viscosity can be estimated as the product of the plateau modulus, $G_1$, and the terminal relaxation time $\tau_{\mathrm{R}}$ (see Eq [14])

$$\eta_{\mathrm{z}} \approx G_1 \tau_{\mathrm{R}} \qquad [22]$$

$\tau_{\mathrm{R}}$ is the longest timescale for polymer chain reconfiguration. For Rouse chains,

$$G_1 = \frac{k_B T c}{N} \quad [23a]$$

$$\tau_R \approx \frac{\eta_s b^3 N^2}{k_B T} \quad [23b]$$

where $c$ is the monomer number density, $b$ is monomer size, $\eta_s$ is the solvent viscosity, and we have neglected numerical prefactors. The resulting zero-shear viscosity follows the scaling relation

$$\eta_z \approx \eta_s N \phi \quad [24a]$$

where $\phi = b^3 c$ is the polymer volume fraction. This relation predicts a linear increase in viscosity with chain length.

Equation [24a] also predicts a linear dependence on $c$, due to the assumed independence between chains. After treating excluded-volume and hydrodynamic interactions, the chain reconfiguration time becomes[158]

$$\tau_R \approx \frac{\eta_s b^3 N^2}{k_B T} \phi^{\frac{2-3\nu}{3\nu-1}} \quad [23b']$$

and the viscosity is now

$$\eta_z \approx \eta_s N \phi^{\frac{1}{3\nu-1}} \quad [24b]$$

where $\nu$ is the Flory exponent. For a theta solvent, $\nu$ = 1/2 and Eq [25b] predicts a quadratic dependence on $\phi$. For associative polymers, breakage of interchain crosslinks becomes the slowest mode of chain reconfiguration. Rubinstein and Semenov[159] introduced a sticker-spacer model for associative polymers, in which stickers, equally spaced along the chain, form crosslinks. The mean lifetime, $\tau_b$, of interchain crosslinks, is determined by their strength, in particular the energy barrier when two stickers separate. If each chain has $f$ stickers, then the chain reconfiguration time is

$$\tau_{\mathrm{R}} = \tau_{\mathrm{b}} f^2 \quad [23b']$$

The viscosity is now

$$\eta_{\mathrm{z}} \approx k_B T \tau_{\mathrm{b}} c f^2 / N \quad [24c]$$

After the breakup of an interchain crosslink between two stickers, they tend to reform the same crosslink as most stickers are engaged in crosslinks. Hence a given crosslink may break up multiple times before one of the stickers finds a new partner, leading to a much longer lifetime than the bare $\tau_{\mathrm{b}}$ and consequently a much larger $\eta_{\mathrm{z}}$.

Condensates can differ greatly in crosslink strength and number of stickers per chain, and thus have widely varying viscosities. This variation is illustrated by the values that we measured for three XXssXX condensates, 0.22, 123, and 856 Pa·s, respectively, for X = M, L, and F.[13] These differences in viscosity correlate with those exhibited by the threshold concentration for phase separation (Figure 4e), but span a much wider range (3856-fold for $\eta_{\mathrm{z}}$ vs. 12.5-fold for $C_{\mathrm{th}}$). The wide variation of $\eta_{\mathrm{z}}$ can be explained by the fact that $C_{\mathrm{th}}$, as an <u>equilibrium</u> property, depends on the well depth $\varepsilon$ of a crosslink, whereas $\tau_{\mathrm{b}}$, as a <u>dynamic</u> property, also depends on an additional activation barrier $\varepsilon_{\mathrm{a}}$ (Figure 5a).[159] That is, $\varepsilon_{\mathrm{a}}$ can produce further variation in $\eta_{\mathrm{z}}$.

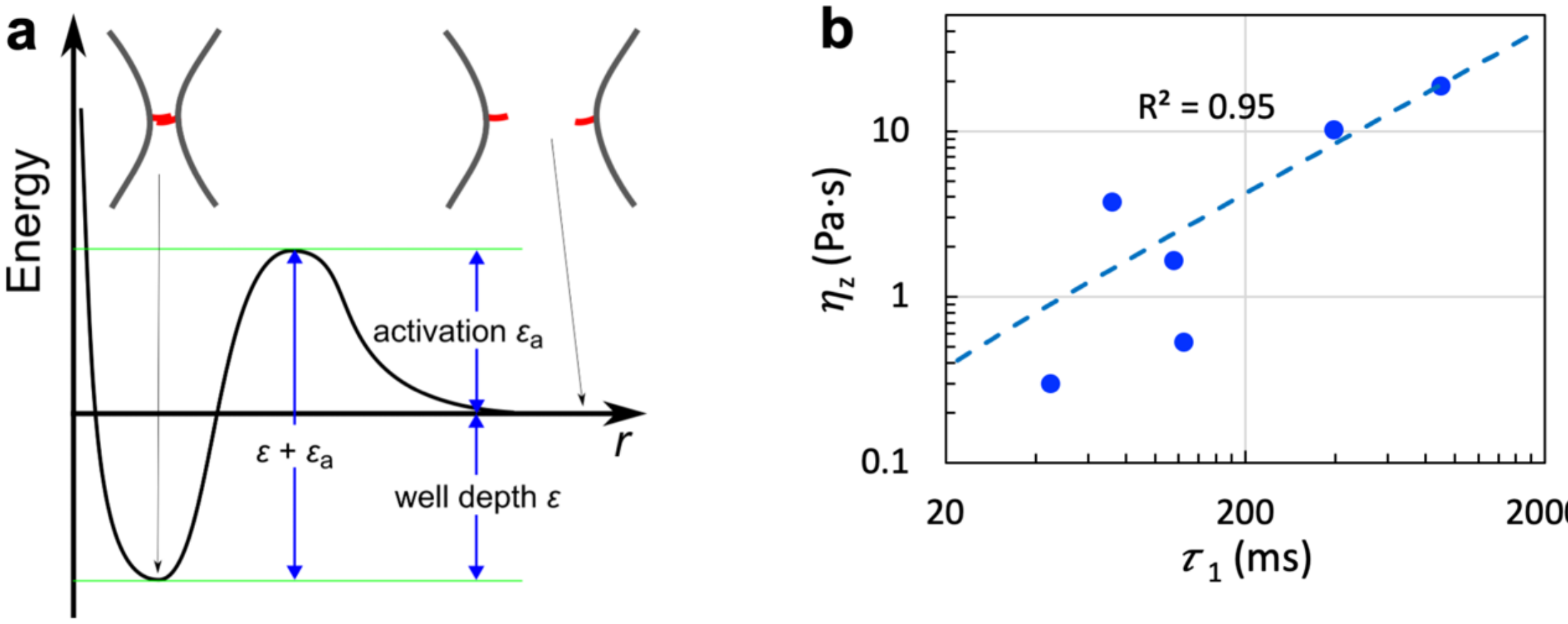

**Figure 5.** Wide variation of $\eta_z$ and correlation between $\eta_z$ and $\tau_1$. (a) Energy surface for the breakup of an interchain crosslink. (b) Correlation between $\eta_z$ and $\tau_1$, with data from refs [90, 100].

To gain molecular insight into viscosity in XXssXX condensates, we tracked the mean-square-displacements (MSDs) of peptide molecules in our all-atom MD simulations (X = L and F; Figure 4e). In the dilute phase, MSDs of both LLssLL and FFssFF were proportional to time and thus conformed to Brownian motion; the diffusion constant from the slope, 10 Å$^2$/ns, was also consistent with expectations for peptides of their size. In the dense phase, the peptides exhibited subdiffusion, which has also been observed experimentally[160] and can be viewed as a manifestation of viscoelasticity.[161] At $t$ = 30 ns, the apparent diffusion coefficients (i.e., MSD/6$t$) were 21 $\times 10^{-3}$ Å$^2$/ns for LLssLL and 9 $\times 10^{-3}$ Å$^2$/ns for FFssFF. The 2.3-fold difference reflects the stronger interactions of FFssFF than LLssLL. The experimental and computational results both demonstrate that interaction strength is the dominant determinant of condensate viscosity, whereas molecular size is not a crucial factor. This point is reinforced by the results on ATP-IDP condensates.[100] ATP, despite its small size, interacts with protamine with high valency (Figure 3e). The zero-shear viscosity, 18.6 Pa·s, is as high as or even higher than those of homotypic or heterotypic IDP condensates.

Equations [14] and [22] suggest that the zero-shear viscosity may be correlated with the stress relaxation time $\tau_1$. Data from six condensates actually show a strong correlation between $\eta_z$ and $\tau_1$ (Figure 5b). Recently, Galvanetto et al.[86] have reported a similar correlation between $\eta_z$ and the chain reconfiguration time $\tau_R$, which was measured by nanosecond fluorescence correlation spectroscopy.

I close this section by noting that, given the relative constancy of surface tension and the wide variation in viscosity, the latter largely accounts for changes in fusion speed. As fusion experiments are relatively simple to perform, one can use the inverse fusion speed ($\tau_{\mathrm{fu}}/R$) as a quick indicator of viscosity magnitude.

## 5. CONCLUDING REMARKS

Based on our studies over the past decade, I have presented here theoretical frameworks for the phase-separation propensities, material states, and material properties, which have largely been missing in the biomolecular condensate literature. Using these frameworks, I have rationalized mechanistic interpretations from our recent experimental and computational studies, and synthesized these studies with prior literature to draw new conclusions.

For phase-separation propensities, I have derived an equation that directly relates the threshold (or saturation) concentration, a widely used metric for phase-separation propensity, to the excess chemical potential in the dense phase, which in turn depends on intermolecular interaction strength and valency. The latter are frequently identified as the drive for phase separation; now there is a firm theoretical basis.

For material states, I posit that they arise from the termination of spinodal decomposition at different stages. Amorphous dense liquids, gels and reversible aggregates, and liquid droplets form upon the termination of spinodal decomposition at the early, middle, and end stages of spinodal decomposition, respectively. Premature termination is due to overly weak intermolecular interactions for amorphous dense liquids and overly strong interactions for aggregates. In addition to high strength, the directionality of interactions also promotes gel formation by favoring tip growth and thereby leading to system-spanning molecular networks. These conclusions are supported by multiple lines of evidence. For example, regarding the order of interaction strength among the four material states, weakening electrostatic attraction by adding salt converts aggregates into droplets.[89] Likewise, upon increasing electrostatic repulsion by lowering pH, the material state of XXssXX peptides changes from droplets or gels to amorphous dense liquids.[13] Moreover, our MD simulations revealed backbone hydrogen bonding as a major determinant of gel formation, with its directionality likely promoting tip growth; the sea-urchin morphology of gels (Figure 2c) is direct evidence of tip growth. I contend that gels and aggregates are different forms of dynamically arrested states, with gels driven by tip growth via directional interactions whereas aggregates driven by monomer addition at interior sites to maximize valency. The relationship between these two material states will require future studies to clarify.

For condensate properties, I have highlighted the crucial roles of the stress relaxation time $\tau_1$, which is determined by the mean lifetime of intermolecular bonds in a condensate. I have used an analytic solution for the shape recovery problem[155] to show that the relative magnitudes of $\tau_1$ and the timescale of the process of interest dictate how the condensate manifests viscoelasticity. If $\tau_1$ is very short, the condensate should appear purely viscous in the process. If $\tau_1$ is short but still close to the timescale of interest, the condensate should exhibit shear thickening; conversely, if $\tau_1$ is very long, then the condensate should exhibit shear thinning. Based on theoretical grounds,[159] I argue that the mean lifetime of intermolecular bonds is responsible for the wide variation in zero-shear viscosity ($\eta_z$) among different condensates. Indeed, I propose a strong correlation between $\eta_z$ and $\tau_1$, which is supported by available data. I also devised a theoretical model to explain the relative constancy of surface tension. Given this constancy, I suggest that the inverse fusion speed can be used as an indicator of viscosity.

Many of the foregoing conclusions are tentative. Validating and expanding these conclusions will further strengthen the theoretical foundations of the biomolecular condensate field. Other opportunities of particular interest include:

*Bridging the gap between all-atom MD simulations and in vitro experiments.* Our all-atom MD simulations of phase separation have significantly enriched our mechanistic understanding of biomolecular condensates. In particular, they revealed backbone hydrogen bonding as a major determinant of gel formation[13] and motivated further experimental tests.[87] Still, these simulations are limited to molecules of relatively small sizes, such as peptides and small IDPs. Moreover, the simulated times are limited to a few μs, far shorter than typical experimental timescales (> 1 ms). Extending the size and time limits, e.g., by taking advantage of deep learning, will tremendously enhance the power of MD simulations in integrating with *in vitro* experiments for deep mechanistic insight.

*Expanding the experimental toolkit for the study of biomolecular condensates.* Optical microscopy and OT are now common techniques for studying biomolecular condensates. Other biophysical techniques, including NMR spectroscopy, dynamic light scattering, FTIR, Raman spectroscopy, and TEM, are gaining use. Still other techniques from materials research, including selected area electron diffraction,[143, 162] field emission scanning electron microscopy,[162] and liquid-phase TEM,[163] can benefit the characterization of biomolecular condensates.

*Bridging the gap between in vitro and in vivo studies.* The *in vivo* function of protamine is to replace histones in the sperm nucleus to achieve extreme chromatin compaction and silence transcription. Our OT-based single-molecule studies have shown that protamine can condense DNA into tangles that are too stable to be unraveled RNA polymerases, leading to the hypothesis that tangle formation is the basis for transcription silencing.[107-109] Ultimately, this hypothesis must be tested by experiments in the sperm nucleus.

## BIOGRAPHY

Huan-Xiang Zhou received his Ph.D. from Drexel University in 1988 and did postdoctoral work at the National Institutes of Health. After faculty appointments at Hong Kong University of Science and Technology, Drexel University, and Florida State University, he moved in 2017 to the University of Illinois Chicago, where he is Professor of Chemistry and Physics and holds an LAS Endowed Chair in the Natural Sciences. He was elected a fellow of the American Association for the Advancement of Science, a fellow of the American Physical Society, and a fellow of the Biophysical Society. His group does theoretical, computational, and experimental research on molecular and cellular biophysics. Current interests include thermodynamic and dynamic properties of biomolecular condensates; membrane association of intrinsically disordered proteins; structures and interactions of *M. tuberculosis* divisome proteins; and functional mechanisms of glutamate-receptor ion channels.

## AKNOWLEDGEMENTS

The conclusions presented in this Account were formed from studies by and discussions with members of my group. I also benefited from discussions with many colleagues, including Jan Spille, Vivek Sharma, Xuezheng Cao, and Zhangli Peng. This work was supported by National Institutes of Health Grant GM118091.

## REFERENCES

(1) Boeynaems, S.; Alberti, S.; Fawzi, N. L.; Mittag, T.; Polymenidou, M.; Rousseau, F.; Schymkowitz, J.; Shorter, J.; Wolozin, B.; Van Den Bosch, L.; et al. Protein Phase Separation: A New Phase in Cell Biology. *Trends Cell Biol* **2018**, *28*, 420-435.

(2) Woodruff, J. B.; Hyman, A. A.; Boke, E. Organization and Function of Non-dynamic Biomolecular Condensates. *Trends Biochem Sci* **2018**, *43*, 81-94.

(3) Alberti, S.; Hyman, A. A. Biomolecular condensates at the nexus of cellular stress, protein aggregation disease and ageing. *Nat Rev Mol Cell Biol* **2021**, *22*, 196-213.

(4) Ranganathan, S.; Liu, J.; Shakhnovich, E. Different states and the associated fates of biomolecular condensates. *Essays Biochem* **2022**, *66*, 849-862.

(5) Wang, Z.; Lou, J.; Zhang, H. Essence determines phenomenon: Assaying the material properties of biological condensates. *J Biol Chem* **2022**, *298*, 101782.

(6) Sanfeliu-Cerdán, N.; Krieg, M. The mechanobiology of biomolecular condensates. *Biophys Rev* **2025**, *6*, 011310.

(7) Mazarakos, K.; Prasad, R.; Zhou, H. X. SpiDec: Computing binodals and interfacial tension of biomolecular condensates from simulations of spinodal decomposition. *Front Mol Biosci* **2022**, *9*, 1021939.

(8) Zhou, H. X.; Kota, D.; Qin, S.; Prasad, R. Fundamental Aspects of Phase-Separated Biomolecular Condensates. *Chem Rev* **2024**, *124*, 8550-8595.

(9) Brangwynne, C. P.; Eckmann, C. R.; Courson, D. S.; Rybarska, A.; Hoege, C.; Gharakhani, J.; Julicher, F.; Hyman, A. A. Germline P Granules Are Liquid Droplets That Localize by Controlled Dissolution/Condensation. *Science* **2009**, *324*, 1729-1732.

(10) Platani, M.; Goldberg, I.; Swedlow, J. R.; Lamond, A. I. In vivo analysis of Cajal body movement, separation, and joining in live human cells. *J Cell Biol* **2000**, *151*, 1561-1574.

(11) Feric, M.; Vaidya, N.; Harmon, T. S.; Mitrea, D. M.; Zhu, L.; Richardson, T. M.; Kriwacki, R. W.; Pappu, R. V.; Brangwynne, C. P. Coexisting Liquid Phases Underlie Nucleolar Subcompartments. *Cell* **2016**, *165*, 1686-1697.

(12) Jain, S.; Wheeler, J. R.; Walters, R. W.; Agrawal, A.; Barsic, A.; Parker, R. ATPase-Modulated Stress Granules Contain a Diverse Proteome and Substructure. *Cell* **2016**, *164*, 487-498.

(13) Zhang, Y.; Prasad, R.; Su, S.; Lee, D.; Zhou, H. X. Amino acid-dependent phase equilibrium and material properties of tetrapeptide condensates. *Cell Rep Phys Sci* **2024**, *5*, 102218.

(14) Cho, W. K.; Spille, J. H.; Hecht, M.; Lee, C.; Li, C.; Grube, V.; Cisse, II. Mediator and RNA Polymerase II Clusters Associate in Transcription-Dependent Condensates. *Science* **2018**, *361*, 412-415.

(15) Cardinaux, F.; Gibaud, T.; Stradner, A.; Schurtenberger, P. Interplay between Spinodal Decomposition and Glass Formation in Proteins Exhibiting Short-Range Attractions. *Phys Rev Lett* **2007**, *99*, 118301.

(16) Lu, P. J.; Zaccarelli, E.; Ciulla, F.; Schofield, A. B.; Sciortino, F.; Weitz, D. A. Gelation of particles with short-range attraction. *Nature* **2008**, *453*, 499-503.

(17) Shin, Y.; Berry, J.; Pannucci, N.; Haataja, M. P.; Toettcher, J. E.; Brangwynne, C. P. Spatiotemporal Control of Intracellular Phase Transitions Using Light-Activated optoDroplets. *Cell* **2017**, *168*, 159-171.e114.

(18) Zhou, P.; Xing, R.; Li, Q.; Li, J.; Yuan, C.; Yan, X. Steering phase-separated droplets to control fibrillar network evolution of supramolecular peptide hydrogels. *Matter* **2023**, *6*, 1945-1963.

(19) Frey, S.; Richter, R. P.; Görlich, D. FG-rich repeats of nuclear pore proteins form a three-dimensional meshwork with hydrogel-like properties. *Science* **2006**, *314*, 815-817.
(20) Putnam, A.; Cassani, M.; Smith, J.; Seydoux, G. A Gel Phase Promotes Condensation of Liquid P Granules in *Caenorhabditis elegans* Embryos. *Nat. Struct. Mol. Biol.* **2019**, *26*, 220-226.
(21) Tanaka, A.; Nakano, T.; Watanabe, K.; Masuda, K.; Honda, G.; Kamata, S.; Yasui, R.; Kozuka-Hata, H.; Watanabe, C.; Chinen, T.; et al. Stress-dependent cell stiffening by tardigrade tolerance proteins that reversibly form a filamentous network and gel. *PLoS Biol* **2022**, *20*, e3001780.
(22) Kato, M.; Han, Tina W.; Xie, S.; Shi, K.; Du, X.; Wu, Leeju C.; Mirzaei, H.; Goldsmith, Elizabeth J.; Longgood, J.; Pei, J.; et al. Cell-free Formation of RNA Granules: Low Complexity Sequence Domains Form Dynamic Fibers within Hydrogels. *Cell* **2012**, *149*, 753-767.
(23) Riback, J. A.; Katanski, C. D.; Kear-Scott, J. L.; Pilipenko, E. V.; Rojek, A. E.; Sosnick, T. R.; Drummond, D. A. Stress-Triggered Phase Separation Is an Adaptive, Evolutionarily Tuned Response. *Cell* **2017**, *168*, 1028-1040.e1019.
(24) Gabryelczyk, B.; Cai, H.; Shi, X.; Sun, Y.; Swinkels, P. J. M.; Salentinig, S.; Pervushin, K.; Miserez, A. Hydrogen bond guidance and aromatic stacking drive liquid-liquid phase separation of intrinsically disordered histidine-rich peptides. *Nat Commun* **2019**, *10*, 5465.
(25) Nie, J.; Zhang, X.; Hu, Z.; Wang, W.; Schroer, M. A.; Ren, J.; Svergun, D.; Chen, A.; Yang, P.; Zeng, A. P. A globular protein exhibits rare phase behavior and forms chemically regulated orthogonal condensates in cells. *Nat Commun* **2025**, *16*, 2449.

(26) Martin, E. W.; Holehouse, A. S.; Peran, I.; Farag, M.; Incicco, J. J.; Bremer, A.; Grace, C. R.; Soranno, A.; Pappu, R. V.; Mittag, T. Valence and Patterning of Aromatic Residues Determine the Phase Behavior of Prion-Like Domains. *Science* **2020**, *367*, 694-699.

(27) Tombs, M. P.; Newsom, B. G.; Wilding, P. Protein solubility: Phase separation in arachin-salt-water systems. *Int J Peptide Protein Res* **1974**, *6*, 253-277.

(28) Tanaka, T.; Ishimoto, C.; Chylack, L. T., Jr. Phase separation of a protein-water mixture in cold cataract in the young rat lens. *Science* **1977**, *197*, 1010-1012.

(29) Taratuta, V. G.; Holschbach, A.; Thurston, G. M.; Blankschtein, D.; Benedek, G. B. Liquid-liquid phase separation of aqueous lysozyme solutions: effects of pH and salt identity. *J Phys Chem* **1990**, *94*, 2140-2144.

(30) Broide, M. L.; Tominc, T. M.; Saxowsky, M. D. Using phase transitions to investigate the effect of salts on protein interactions. *Phys Rev E* **1996**, *53*, 6325-6335.

(31) Muschol, M.; Rosenberger, F. Liquid–Liquid Phase Separation in Supersaturated Lysozyme Solutions and Associated Precipitate Formation/Crystallization. *J. Chem. Phys.* **1997**, *107*, 1953-1962.

(32) Manno, M.; Xiao, C.; Bulone, D.; Martorana, V.; San Biagio, P. L. Thermodynamic instability in supersaturated lysozyme solutions: Effect of salt and role of concentration fluctuations. *Phys Rev E* **2003**, *68*, 011904.

(33) Petsev, D. N.; Wu, X.; Galkin, O.; Vekilov, P. G. Thermodynamic Functions of Concentrated Protein Solutions from Phase Equilibria. *J Phys Chem B* **2003**, *107*, 3921-3926.

(34) Wang, Y.; Annunziata, O. Liquid−Liquid Phase Transition of Protein Aqueous Solutions Isothermally Induced by Protein Cross-Linking. *Langmuir* **2008**, *24*, 2799-2807.

(35) Siezen, R. J.; Fisch, M. R.; Slingsby, C.; Benedek, G. B. Opacification of gamma-crystallin solutions from calf lens in relation to cold cataract formation. *Proc Natl Acad Sci U S A* **1985**, *82*, 1701-1705.

(36) Thomson, J. A.; Schurtenberger, P.; Thurston, G. M.; Benedek, G. B. Binary liquid phase separation and critical phenomena in a protein/water solution. *Proc Natl Acad Sci U S A* **1987**, *84*, 7079-7083.

(37) Broide, M. L.; Berland, C. R.; Pande, J.; Ogun, O. O.; Benedek, G. B. Binary-liquid phase separation of lens protein solutions. *Proc Natl Acad Sci U S A* **1991**, *88*, 5660-5664.

(38) Annunziata, O.; Pande, A.; Pande, J.; Ogun, O.; Lubsen, N. H.; Benedek, G. B. Oligomerization and phase transitions in aqueous solutions of native and truncated human beta B1-crystallin. *Biochemistry* **2005**, *44*, 1316-1328.

(39) Zhang, F.; Skoda, M. W.; Jacobs, R. M.; Zorn, S.; Martin, R. A.; Martin, C. M.; Clark, G. F.; Weggler, S.; Hildebrandt, A.; Kohlbacher, O.; et al. Reentrant condensation of proteins in solution induced by multivalent counterions. *Phys Rev Lett* **2008**, *101*, 148101.

(40) Dumetz, A. C.; Chockla, A. M.; Kaler, E. W.; Lenhoff, A. M. Protein Phase Behavior in Aqueous Solutions: Crystallization, Liquid-Liquid Phase Separation, Gels, and Aggregates. *Biophys J* **2008**, *94*, 570-583.

(41) Hansen, J.; Moll, C. J.; López Flores, L.; Castañeda-Priego, R.; Medina-Noyola, M.; Egelhaaf, S. U.; Platten, F. Phase separation and dynamical arrest of protein solutions dominated by short-range attractions. *J Chem Phys* **2023**, *158*, 024904.

(42) Andrews, T. XVIII. The Bakerian Lecture.—On the continuity of the gaseous and liquid states of matter. *Phil Trans R Soc London* **1869**, *159*, 575-590.

(43) Alexejew, W. Ueber Lösungen (About solutions). *Ann Phys Chem* **1886**, *Ser 3 V 28*, 305–338.

(44) Beijerinck, M. Über eine Eigentümlichkeit der löslichen Stärke (About a peculiarity of soluble starch). *Zentralbl Bakteriol* **1896**, *2*, 679-699.

(45) Tiebackx, F. W. Gleichzeitige Ausflockung zweier Kolloide (Simultaneous flocculation of two colloids). *Z. Chem. Ind. Kolloide* **1911**, *8*, 198-201.

(46) Bungenberg de Jong, H. G.; Kruyt, H. R. Coacervation (partial miscibility in colloid systems). *Proc Koninkl Med Akad Wetershap* **1929**, *32*, 849-856.

(47) Brönsted, J. N.; Volqvartz, K. Solubility and swelling of high polymers. *Trans Faraday Soc* **1939**, *35*, 576-579.

(48) Brønsted, J. N.; Volqvartz, K. Solubility and swelling of high polymers in ternary mixtures. *Trans Faraday Soc* **1940**, *35*, 619-624.

(49) Tanner, L. E.; Ray, R. Phase separation in Zr-Ti-Be metallic glasses. *Scr Metall* **1980**, *14*, 657-662.

(50) Miller, M. K.; Bentley, J. Microstructural characterization of primary coolant pipe steel. *J. Phys. Colloques* **1986**, *47*, C7 239-244.

(51) Busch, R.; Schneider, S.; Peker, A.; Johnson, W. L. Decomposition and primary crystallization in undercooled $Zr_{41.2}Ti_{13.8}Cu_{12.5}Ni_{10.0}Be_{22.5}$ melts. *Appl Phys Lett* **1995**, *67*, 1544-1546.

(52) Van der Waals, J. D. Over de Continuïteit van den Gas- en Vloeistoftoestand (On the Continuity of the Gaseous and Liquid State). University of Leiden, 1873.

(53) Huggins, M. L. Solutions of Long Chain Compounds. *J Chem Phys* **1941**, *9*, 440-440.

(54) Flory, P. J. Thermodynamics of High Polymer Solutions. *J Chem Phys* **1942**, *10*, 51-61.

(55) Gibbs, J. W. On the Equilibrium of Heterogeneous Substances. *Trans Conn Acad Arts Sci* **1876; 1878**, *3*, 108–248; 343–524.

(56) Mazarakos, K.; Zhou, H. X. Multiphase organization is a second phase transition within multi-component biomolecular condensates. *J Chem Phys* **2022**, *156*, 191104.

(57) Cahn, J. W. Phase Separation by Spinodal Decomposition in Isotropic Systems. *J Chem Phys* **1965**, *42*, 93-99.

(58) Siggia, E. D. Late stages of spinodal decomposition in binary mixtures. *Phys Rev A* **1979**, *20*, 595-605.

(59) Dhont, J. K. G. Spinodal decomposition of colloids in the initial and intermediate stages. *J Chem Phys* **1996**, *105*, 5112-5125.

(60) Rao, M.; Levesque, D. Surface-Structure of a Liquid-Film. *J. Chem. Phys.* **1976**, *65*, 3233-3236.

(61) Panagiotopoulos, A. Z. Direct Determination of Phase Coexistence Properties of Fluids by Monte Carlo Simulations in a New Ensemble. *Mol. Phys.* **1987**, *61*, 813-826.

(62) Qin, S.; Zhou, H. X. Fast Method for Computing Chemical Potentials and Liquid-Liquid Phase Equilibria of Macromolecular Solutions. *J. Phys. Chem. B* **2016**, *120*, 8164-8174.

(63) McCarty, J.; Delaney, K. T.; Danielsen, S. P. O.; Fredrickson, G. H.; Shea, J. E. Complete Phase Diagram for Liquid-Liquid Phase Separation of Intrinsically Disordered Proteins. *J. Phys. Chem. Lett.* **2019**, *10*, 1644-1652.

(64) Laghmach, R.; Potoyan, D. A. Liquid–liquid phase separation driven compartmentalization of reactive nucleoplasm. *Phys Biol* **2021**, *18*, 015001.

(65) Lin, Y. H.; Wessén, J.; Pal, T.; Das, S.; Chan, H. S. Numerical Techniques for Applications of Analytical Theories to Sequence-Dependent Phase Separations of Intrinsically Disordered Proteins. *Methods Mol Biol* **2023**, *2563*, 51-94.

(66) Mazarakos, K.; Qin, S.; Zhou, H. X. Calculating Binodals and Interfacial Tension of Phase-Separated Condensates from Molecular Simulations with Finite-Size Corrections. *Methods Mol. Biol.* **2023**, *2563*, 1-35.

(67) Groves, S. M.; Lu, M. J.; Alvarez-Yela, A. C.; Gerardo-Ramírez, M.; Stukenberg, P. T.; Lowengrub, J. S.; Janes, K. A. Cahn-Hilliard dynamical models for condensed biomolecular systems. *bioRxiv* **2025**.

(68) Makasewicz, K.; Schneider, T. N.; Mathur, P.; Stavrakis, S.; deMello, A. J.; Arosio, P. Formation of multicompartment structures through aging of protein-RNA condensates. *Biophys J* **2025**, *124*, 115-124.

(69) Qin, S.; Zhou, H. X. A Fixed-Volume Variant of Gibbs-Ensemble Monte Carlo Yields Significant Speedup in Binodal Calculation. *arXiv* **2025**, 2512.18899

(70) Lomakin, A.; Asherie, N.; Benedek, G. B. Monte Carlo study of phase separation in aqueous protein solutions. *J Chem Phys* **1996**, *104*, 1646-1656.

(71) Das, S.; Amin, A. N.; Lin, Y. H.; Chan, H. S. Coarse-Grained Residue-Based Models of Disordered Protein Condensates: Utility and Limitations of Simple Charge Pattern Parameters. *Phys. Chem. Chem. Phys.* **2018**, *20*, 28558-28574.

(72) Dignon, G. L.; Zheng, W.; Kim, Y. C.; Best, R. B.; Mittal, J. Sequence Determinants of Protein Phase Behavior from a Coarse-Grained Model. *PLoS Comput. Biol.* **2018**, *14*, e1005941.

(73) Nguemaha, V.; Zhou, H. X. Liquid-Liquid Phase Separation of Patchy Particles Illuminates Diverse Effects of Regulatory Components on Protein Droplet Formation. *Sci. Rep.* **2018**, *8*, 6728.

(74) Boeynaems, S.; Holehouse, A. S.; Weinhardt, V.; Kovacs, D.; Van Lindt, J.; Larabell, C.; Van Den Bosch, L.; Das, R.; Tompa, P. S.; Pappu, R. V.; et al. Spontaneous Driving Forces Give Rise to Protein-RNA Condensates with Coexisting Phases and Complex Material Properties. *Proc. Natl. Acad. Sci. USA* **2019**, *116*, 7889-7898.

(75) Espinosa, J. R.; Joseph, J. A.; Sanchez-Burgos, I.; Garaizar, A.; Frenkel, D.; Collepardo-Guevara, R. Liquid Network Connectivity Regulates the Stability and Composition of Biomolecular Condensates with Many Components. *Proc. Natl. Acad. Sci. USA* **2020**, *117*, 13238-13247.

(76) Benayad, Z.; von Bulow, S.; Stelzl, L. S.; Hummer, G. Simulation of FUS Protein Condensates with an Adapted Coarse-Grained Model. *J. Chem. Theory Comput.* **2021**, *17*, 525-537.

(77) Joseph, J. A.; Reinhardt, A.; Aguirre, A.; Chew, P. Y.; Russell, K. O.; Espinosa, J. R.; Garaizar, A.; Collepardo-Guevara, R. Physics-Driven Coarse-Grained Model for Biomolecular Phase Separation with Near-Quantitative Accuracy. *Nat. Comput. Sci.* **2021**, *1*, 732-743.

(78) Mazarakos, K.; Zhou, H. X. Macromolecular regulators have matching effects on the phase equilibrium and interfacial tension of biomolecular condensates. *Protein Sci* **2021**, *30*, 1360-1370.

(79) Galvanetto, N.; Ivanović, M. T.; Chowdhury, A.; Sottini, A.; Nüesch, M. F.; Nettels, D.; Best, R. B.; Schuler, B. Extreme Dynamics in a Biomolecular Condensate. *Nature* **2023**, *619*, 876-883.

(80) Liu, S.; Wang, C.; Latham, A. P.; Ding, X.; Zhang, B. OpenABC Enables Flexible, Simplified, and Efficient GPU Accelerated Simulations of Biomolecular Condensates. *PLoS Comput. Biol.* **2023**, *19*, e1011442.

(81) Qin, S.; Zhou, H. X. Atomistic Modeling of Liquid-Liquid Phase Equilibrium Explains Dependence of Critical Temperature on gamma-Crystallin Sequence. *Commun. Biol.* **2023**, *6*, 886.

(82) Tesei, G.; Lindorff-Larsen, K. Improved Predictions of Phase Behaviour of Intrinsically Disordered Proteins by Tuning the Interaction Range. *Open Res. Eur.* **2023**, *2*, 94.

(83) An, Y.; Webb, M. A.; Jacobs, W. M. Active Learning of the Thermodynamics-Dynamics Trade-off in Protein Condensates. *Sci. Adv.* **2024**, *10*, eadj2448.

(84) MacAinsh, M.; Dey, S.; Zhou, H. X. Direct and indirect salt effects on homotypic phase separation. *eLife* **2024**, *13*, RP100282.

(85) Zhang, Y.; Li, S.; Gong, X.; Chen, J. Toward Accurate Simulation of Coupling between Protein Secondary Structure and Phase Separation. *J. Am. Chem. Soc.* **2024**, *146*, 342-357.

(86) Galvanetto, N.; Ivanović, M. T.; Del Grosso, S. A.; Chowdhury, A.; Sottini, A.; Nettels, D.; Best, R. B.; Schuler, B. Material properties of biomolecular condensates emerge from nanoscale dynamics. *Proc Natl Acad Sci U S A* **2025**, *122*, e2424135122.

(87) Zhang, Y.; Prasad, R.; Zhou, H.-X. Backbone Hydrogen Bonding as a Determinant of Condensate Material States. *bioRxiv* **2025**, 2025.2012.2021.695814.

(88) Muhammedkutty, F. N. K.; MacAinsh, M.; Zhou, H. X. Atomistic molecular dynamics simulations of intrinsically disordered proteins. *Curr Opin Struct Biol* **2025**, *92*, 103029.

(89) Ghosh, A.; Zhou, H. X. Determinants for Fusion Speed of Biomolecular Droplets. *Angew. Chem. Int. Ed.* **2020**, *59*, 20837-20840.

(90) Ghosh, A.; Kota, D.; Zhou, H. X. Shear Relaxation Governs Fusion Dynamics of Biomolecular Condensates. *Nat. Commun.* **2021**, *12*, 5995.

(91) Kota, D.; Zhou, H. X. Macromolecular Regulation of the Material Properties of Biomolecular Condensates. *J Phys Chem Lett* **2022**, 5285-5290.

(92) Qin, S.; Zhou, H. X. Calculation of Second Virial Coefficients of Atomistic Proteins Using Fast Fourier Transform. *J Phys Chem B* **2019**, *123*, 8203-8215.

(93) Ahn, S. H.; Qin, S.; Zhang, J. Z.; McCammon, J. A.; Zhang, J.; Zhou, H. X. Characterizing protein kinase A (PKA) subunits as macromolecular regulators of PKA RIalpha liquid-liquid phase separation. *J Chem Phys* **2021**, *154*, 221101.

(94) Jia, T. Z.; Hentrich, C.; Szostak, J. W. Rapid RNA Exchange in Aqueous Two-Phase System and Coacervate Droplets. *Orig. Life Evol. Biosph.* **2014**, *44*, 1-12.

(95) Kang, J.; Lim, L.; Song, J. ATP Enhances at Low Concentrations but Dissolves at High Concentrations Liquid-Liquid Phase Separation (LLPS) of ALS/FTD-Causing FUS. *Biochem. Biophys. Res. Commun.* **2018**, *504*, 545-551.

(96) Nakashima, K. K.; Baaij, J. F.; Spruijt, E. Reversible Generation of Coacervate Droplets in an Enzymatic Network. *Soft Matter* **2018**, *14*, 361-367.

(97) Babinchak, W. M.; Dumm, B. K.; Venus, S.; Boyko, S.; Putnam, A. A.; Jankowsky, E.; Surewicz, W. K. Small Molecules as Potent Biphasic Modulators of Protein Liquid-Liquid Phase Separation. *Nat. Commun.* **2020**, *11*, 5574.

(98) Dang, M.; Lim, L.; Kang, J.; Song, J. ATP Biphasically Modulates LLPS of TDP-43 PLD by Specifically Binding Arginine Residues. *Commun. Biol.* **2021**, *4*, 714.

(99) Toyama, Y.; Rangadurai, A. K.; Forman-Kay, J. D.; Kay, L. E. Mapping the Per-Residue Surface Electrostatic Potential of CAPRIN1 along Its Phase-Separation Trajectory. *Proc. Natl. Acad. Sci. USA* **2022**, *119*, e2210492119.

(100) Kota, D.; Prasad, R.; Zhou, H. X. Adenosine Triphosphate Mediates Phase Separation of Disordered Basic Proteins by Bridging Intermolecular Interaction Networks. *J. Am. Chem. Soc.* **2024**, *146*, 1326-1336.

(101) Abbas, M.; Lipiński, W. P.; Nakashima, K. K.; Huck, W. T. S.; Spruijt, E. A Short Peptide Synthon for Liquid–Liquid Phase Separation. *Nat Chem* **2021**, *13*, 1046-1054.

(102) Abbas, M.; Law, J. O.; Grellscheid, S. N.; Huck, W. T. S.; Spruijt, E. Peptide-Based Coacervate-Core Vesicles with Semipermeable Membranes. *Adv Mater* **2022**, *34*, 2202913.

(103) Kitaoka, M.; Yamashita, Y. M. Running the gauntlet: challenges to genome integrity in spermiogenesis. *Nucleus* **2024**, *15*, 2339220.

(104) Moritz, L.; Hammoud, S. S. The Art of Packaging the Sperm Genome: Molecular and Structural Basis of the Histone-To-Protamine Exchange. *Front Endocrinol* **2022**, *13*, 895502.

(105) Rathke, C.; Baarends, W. M.; Awe, S.; Renkawitz-Pohl, R. Chromatin dynamics during spermiogenesis. *Biochim Biophys Acta* **2014**, *1839*, 155-168.

(106) Ward, W. S.; Coffey, D. S. DNA packaging and organization in mammalian spermatozoa: comparison with somatic cells. *Biol Reprod* **1991**, *44*, 569-574.

(107) Ahlawat, V.; Dhiman, A.; Mudiyanselage, H. E.; Zhou, H. X. Protamine-Mediated Tangles Produce Extreme Deoxyribonucleic Acid Compaction. *J Am Chem Soc* **2024**, *146*, 30668-30677.

(108) Ahlawat, V.; Mudiyanselage, H. E.; Kota, D.; Zhou, H. X. Measuring bridging forces in protein-DNA condensates. *Proc Natl Acad Sci U S A* **2026**, *123*, e2523058123.

(109) Ahlawat, V.; Kota, D.; Zhou, H.-X. Counteraction of HMGB1 at ss-dsDNA junctions maintains liquidity of protamine-DNA co-condensates. *bioRxiv* **2026**, 2026.2002.2024.707832.

(110) Kim, J.; Qin, S.; Zhou, H. X.; Rosen, M. K. Surface Charge Can Modulate Phase Separation of Multidomain Proteins. *J. Am. Chem. Soc.* **2024**, *146*, 3383-3395.

(111) Widom, B. Some Topics in the Theory of Fluids. *J Chem Phys* **1963**, *39*, 2808-2812.

(112) Deng, Y.; Roux, B. Hydration of Amino Acid Side Chains:  Nonpolar and Electrostatic Contributions Calculated from Staged Molecular Dynamics Free Energy Simulations with Explicit Water Molecules. *J Phys Chem B* **2004**, *108*, 16567-16576.

(113) Shirts, M. R.; Pande, V. S. Solvation free energies of amino acid side chain analogs for common molecular mechanics water models. *J Chem Phys* **2005**, *122*, 134508.

(114) Wang, J.; Choi, J. M.; Holehouse, A. S.; Lee, H. O.; Zhang, X.; Jahnel, M.; Maharana, S.; Lemaitre, R.; Pozniakovsky, A.; Drechsel, D.; et al. A Molecular Grammar Governing the Driving Forces for Phase Separation of Prion-like RNA Binding Proteins. *Cell* **2018**, *174*, 688-699 e616.

(115) Dzuricky, M.; Rogers, B. A.; Shahid, A.; Cremer, P. S.; Chilkoti, A. De novo Engineering of Intracellular Condensates Using Artificial Disordered Proteins. *Nat. Chem.* **2020**, *12*, 814-825.

(116) Schuster, B. S.; Dignon, G. L.; Tang, W. S.; Kelley, F. M.; Ranganath, A. K.; Jahnke, C. N.; Simpkins, A. G.; Regy, R. M.; Hammer, D. A.; Good, M. C.; et al. Identifying Sequence Perturbations to an Intrinsically Disordered Protein That Determine Its Phase-Separation Behavior. *Proc. Natl. Acad. Sci. USA* **2020**, *117*, 11421-11431.

(117) Wong, L. E.; Kim, T. H.; Muhandiram, D. R.; Forman-Kay, J. D.; Kay, L. E. NMR Experiments for Studies of Dilute and Condensed Protein Phases: Application to the Phase-Separating Protein CAPRIN1. *J. Am. Chem. Soc.* **2020**, *142*, 2471-2489.

(118) Bremer, A.; Farag, M.; Borcherds, W. M.; Peran, I.; Martin, E. W.; Pappu, R. V.; Mittag, T. Deciphering How Naturally Occurring Sequence Features Impact the Phase Behaviours of Disordered Prion-like Domains. *Nat. Chem.* **2022**, *14*, 196-207.

(119) Qin, S.; Zhou, H. X. Predicting the sequence-dependent backbone dynamics of intrinsically disordered proteins. *eLife* **2024**, *12*, RP 88958.

(120) Kim, T. H.; Payliss, B. J.; Nosella, M. L.; Lee, I. T. W.; Toyama, Y.; Forman-Kay, J. D.; Kay, L. E. Interaction hot spots for phase separation revealed by NMR studies of a CAPRIN1 condensed phase. *Proc Natl Acad Sci U S A* **2021**, *118*, e2104897118.

(121) Ginell, G. M.; Emenecker, R. J.; Lotthammer, J. M.; Keeley, A. T.; Plassmeyer, S. P.; Razo, N.; Usher, E. T.; Pelham, J. F.; Holehouse, A. S. Sequence-based prediction of intermolecular interactions driven by disordered regions. *Science* **2025**, *388*, eadq8381.

(122) Zhou, H. X. Correlated Segments of Intrinsically Disordered Proteins as Drivers of Homotypic Phase Separation. *JACS Au* **2025**, *5*, 4361-4369.

(123) Berry, J.; Weber, S. C.; Vaidya, N.; Haataja, M.; Brangwynne, C. P. RNA transcription modulates phase transition-driven nuclear body assembly. *Proc Natl Acad Sci U S A* **2015**, *112*, E5237-5245.

(124) Burke, K. A.; Janke, A. M.; Rhine, C. L.; Fawzi, N. L. Residue-by-Residue View of In Vitro FUS Granules that Bind the C-Terminal Domain of RNA Polymerase II. *Mol Cell* **2015**, *60*, 231-241.

(125) Zhang, H.; Elbaum-Garfinkle, S.; Langdon, E. M.; Taylor, N.; Occhipinti, P.; Bridges, A. A.; Brangwynne, C. P.; Gladfelter, A. S. RNA Controls PolyQ Protein Phase Transitions. *Mol Cell* **2015**, *60*, 220-230.

(126) Brady, J. P.; Farber, P. J.; Sekhar, A.; Lin, Y. H.; Huang, R.; Bah, A.; Nott, T. J.; Chan, H. S.; Baldwin, A. J.; Forman-Kay, J. D.; et al. Structural and hydrodynamic properties of an intrinsically disordered region of a germ cell-specific protein on phase separation. *Proc Natl Acad Sci U S A* **2017**, *114*, E8194-e8203.

(127) Kim, S.; Yoo, H. Y.; Huang, J.; Lee, Y.; Park, S.; Park, Y.; Jin, S.; Jung, Y. M.; Zeng, H.; Hwang, D. S.; et al. Salt Triggers the Simple Coacervation of an Underwater Adhesive When Cations Meet Aromatic pi Electrons in Seawater. *ACS Nano* **2017**, *11*, 6764-6772.

(128) Strom, A. R.; Emelyanov, A. V.; Mir, M.; Fyodorov, D. V.; Darzacq, X.; Karpen, G. H. Phase separation drives heterochromatin domain formation. *Nature* **2017**, *547*, 241-245.

(129) Wei, M. T.; Elbaum-Garfinkle, S.; Holehouse, A. S.; Chen, C. C.; Feric, M.; Arnold, C. B.; Priestley, R. D.; Pappu, R. V.; Brangwynne, C. P. Phase behaviour of disordered proteins underlying low density and high permeability of liquid organelles. *Nat Chem* **2017**, *9*, 1118-1125.

(130) Dao, T. P.; Kolaitis, R. M.; Kim, H. J.; O'Donovan, K.; Martyniak, B.; Colicino, E.; Hehnly, H.; Taylor, J. P.; Castaneda, C. A. Ubiquitin Modulates Liquid-Liquid Phase Separation of UBQLN2 via Disruption of Multivalent Interactions. *Mol Cell* **2018**, *69*, 965-978 e966.

(131) Reed, E. H.; Hammer, D. A. Redox sensitive protein droplets from recombinant oleosin. *Soft Matter* **2018**, *14*, 6506-6513.

(132) Babinchak, W. M.; Haider, R.; Dumm, B. K.; Sarkar, P.; Surewicz, K.; Choi, J. K.; Surewicz, W. K. The role of liquid-liquid phase separation in aggregation of the TDP-43 low-complexity domain. *J Biol Chem* **2019**, *294*, 6306-6317.

(133) Le Ferrand, H.; Duchamp, M.; Gabryelczyk, B.; Cai, H.; Miserez, A. Time-Resolved Observations of Liquid-Liquid Phase Separation at the Nanoscale Using in Situ Liquid Transmission Electron Microscopy. *J Am Chem Soc* **2019**, *141*, 7202-7210.

(134) Tsang, B.; Arsenault, J.; Vernon, R. M.; Lin, H.; Sonenberg, N.; Wang, L.-Y.; Bah, A.; Forman-Kay, J. D. Phosphoregulated FMRP Phase Separation Models Activity-Dependent Translation through Bidirectional Control of mRNA Granule Formation. *Proc. Natl. Acad. Sci. USA* **2019**, *116*, 4218-4227.

(135) Agarwal, A.; Rai, S. K.; Avni, A.; Mukhopadhyay, S. An intrinsically disordered pathological prion variant Y145Stop converts into self-seeding amyloids via liquid-liquid phase separation. *Proc Natl Acad Sci U S A* **2021**, *118*, e2100968118.

(136) Krainer, G.; Welsh, T. J.; Joseph, J. A.; Espinosa, J. R.; Wittmann, S.; de Csillery, E.; Sridhar, A.; Toprakcioglu, Z.; Gudiskyte, G.; Czekalska, M. A.; et al. Reentrant liquid condensate phase of proteins is stabilized by hydrophobic and non-ionic interactions. *Nat Commun* **2021**, *12*, 1085.

(137) Martin, E. W.; Thomasen, F. E.; Milkovic, N. M.; Cuneo, M. J.; Grace, C. R.; Nourse, A.; Lindorff-Larsen, K.; Mittag, T. Interplay of folded domains and the disordered low-complexity domain in mediating hnRNPA1 phase separation. *Nucleic Acids Res* **2021**, *49*, 2931-2945.

(138) Otis, J. B.; Sharpe, S. Sequence Context and Complex Hofmeister Salt Interactions Dictate Phase Separation Propensity of Resilin-like Polypeptides. *Biomacromolecules* **2022**, *23*, 5225-5238.

(139) Lin, Y. H.; Kim, T. H.; Das, S.; Pal, T.; Wessén, J.; Rangadurai, A. K.; Kay, L. E.; Forman-Kay, J. D.; Chan, H. S. Electrostatics of salt-dependent reentrant phase behaviors highlights diverse roles of ATP in biomolecular condensates. *eLife* **2025**, *13*, RP100284.

(140) Liu, C.; Pande, J.; Lomakin, A.; Ogun, O.; Benedek, G. B. Aggregation in aqueous solutions of bovine lens gamma-crystallins: special role of gamma(s). *Invest Ophthalmol Vis Sci* **1998**, *39*, 1609-1619.

(141) Pande, J.; Lomakin, A.; Fine, B.; Ogun, O.; Sokolinski, I.; Benedek, G. Oxidation of gamma II-crystallin solutions yields dimers with a high phase separation temperature. *Proc Natl Acad Sci U S A* **1995**, *92*, 1067-1071.

(142) Asherie, N.; Pande, J.; Lomakin, A.; Ogun, O.; Hanson, S. R.; Smith, J. B.; Benedek, G. B. Oligomerization and phase separation in globular protein solutions. *Biophys Chem* **1998**, *75*, 213-227.

(143) Smeets, P. J. M.; Finney, A. R.; Habraken, W. J. E. M.; Nudelman, F.; Friedrich, H.; Laven, J.; De Yoreo, J. J.; Rodger, P. M.; Sommerdijk, N. A. J. M. A classical view on nonclassical nucleation. *Proc Natl Acad Sci U S A* **2017**, *114*, E7882-E7890.

(144) Hohenberg, P. C.; Halperin, B. I. Theory of dynamic critical phenomena. *Rev Mod Phys* **1977**, *49*, 435-479.

(145) Kallala, M.; Jullien, R.; Cabane, B. Crossover from gelation to precipitation. *J Phys II* **1992**, *2*, 7-25.

(146) Zhou, H. X. Determination of Condensate Material Properties from Droplet Deformation. *J. Phys. Chem. B* **2020**, *124*, 8372-8379.

(147) Fowler, R. H. A tentative statistical theory of Macleod's equation for surface tension, and the parachor. *Proc. Roy. Soc. (London)* **1937**, *A159*, 229-246.

(148) Kirkwood, J. G.; Buff, F. P. The Statistical Mechanical Theory of Surface Tension. *J. Chem. Phys.* **1949**, *17*, 338-343.

(149) Aarts, D. G.; Schmidt, M.; Lekkerkerker, H. N. Direct visual observation of thermal capillary waves. *Science* **2004**, *304*, 847-850.

(150) Alshareedah, I.; Moosa, M. M.; Pham, M.; Potoyan, D. A.; Banerjee, P. R. Programmable Viscoelasticity in Protein-RNA Condensates with Disordered Sticker-Spacer Polypeptides. *Nat. Commun.* **2021**, *12*, 6620.

(151) Zhou, H. X. Viscoelasticity of Biomolecular Condensates Conforms to the Jeffreys Model. *J. Chem. Phys.* **2021**, *154*, 041103.

(152) Cox, W. P.; Merz, E. H. Correlation of dynamic and steady flow viscosities. *J Polym Sci* **1958**, *28*, 619-622.

(153) Cross, M. M. Rheology of non-Newtonian fluids: A new flow equation for pseudoplastic systems. *J Colloid Sci* **1965**, *20*, 417-437.

(154) Cross, M. M. Relation between viscoelasticity and shear-thinning behaviour in liquids. *Rheol Acta* **1979**, *18*, 609-614.

(155) Zhou, H. X. Shape recovery of deformed biomolecular droplets: Dependence on condensate viscoelasticity. *J Chem Phys* **2021**, *155*, 145102.

(156) Naderi, M. M.; Peng, Z.; Zhou, H.-X. Fusion Dynamics of Viscoelastic Droplets: Similarities and Differences to Shape Recovery. *Biophys J* **2026**, submitted.

(157) Doi, M.; Edwards, S. F. *The Theory of Polymer Dynamics*; Clarendon Press, 1988.

(158) Rubinstein, M.; Colby, R. H. *Polymer Physics*; Oxford University Press, 2003.

(159) Rubinstein, M.; Semenov, A. N. Thermoreversible Gelation in Solutions of Associating Polymers. 2. Linear Dynamics. *Macromolecules* **1998**, *31*, 1386-1397.

(160) Shen, Z.; Jia, B.; Xu, Y.; Wessén, J.; Pal, T.; Chan, H. S.; Du, S.; Zhang, M. Biological Condensates Form Percolated Networks with Molecular Motion Properties Distinctly Different from Dilute Solutions. *eLife* **2023**, *12*, e81907.

(161) Grimm, M.; Jeney, S.; Franosch, T. Brownian motion in a Maxwell fluid. *Soft Matter* **2011**, *7*, 2076-2084.

(162) Hiew, S. H.; Lu, Y.; Han, H.; Gonçalves, R. A.; Alfarano, S. R.; Mezzenga, R.; Parikh, A. N.; Mu, Y.; Miserez, A. Modulation of Mechanical Properties of Short Bioinspired Peptide Materials by Single Amino-Acid Mutations. *J Am Chem Soc* **2023**, *145*, 3382-3393.

(163) Raimbault, J.; Chevallard, C.; Ihiawakrim, D.; Ramnarain, V.; Ersen, O.; Gobeaux, F.; Carriere, D. Dense Liquid Precursor in Mineral Crystallization: Spinodal Morphology and High Viscosity Evidenced by Electron Imaging. *Nano Lett* **2025**, *25*, 2275-2282.